\documentclass[12pt,a4paper]{article}
\pdfoutput=1
\usepackage{jheppub}

\usepackage{bm}
\usepackage{amsmath}
\usepackage{amssymb}
\usepackage{ifsym}
\usepackage{amsthm}
\usepackage{amsfonts}
\usepackage{ccaption}
\usepackage{booktabs}
\usepackage{mathrsfs}

\makeatletter
\@addtoreset{equation}{section}

\makeatletter
\renewcommand\section{\@startsection {section}{1}{\z@}%
                                   {-3.5ex \@plus -1ex \@minus -.2ex}%nn
                                   {2.3ex \@plus.2ex}%
                                   {\normalfont\large\bfseries}}
\renewcommand\subsection{\@startsection{subsection}{2}{\z@}%
                                     {-3.25ex\@plus -1ex \@minus -.2ex}%
                                     {1.5ex \@plus .2ex}%
                                     {\normalfont\bfseries}}

  \captionnamefont{\bfseries}
  \captiontitlefont{\small\sffamily}
  \captiondelim{: }
  \hangcaption

\def\sec#1{\S\ref{#1}}
\def\fig#1{Fig.\,\ref{#1}}
\def\req#1{(\ref{#1})}

\definecolor{rust}{rgb}{0.8,0.2,0.2}

\def\GCW{strip wedge}
\def\GEW{rim wedge}

\def\bulk{{\cal M}}
\def\bbulk{\bar {\cal M}}
\def\bdy{\partial {\cal M}}  % alternately {{\cal N}} or {{\dot \bulk}}

\def\Jbdy{J_{\partial}}
\def\Ibdy{I_{\partial}}
\def\Dbdy{D_{\partial}}

\def\REbdy{E_{\ts}}  %boundary residual entropy
\def\REbulk{E_{\Dhole}}  %bulk residual entropy

\def\ts{{\cal T}}			% bdy time strip 
\def\tsend{{\Sigma}}		% spacelike ends of bdy strip 
\def\tsbulk{{\cal T}_{\bcrv}}
\def\tsbulkend{{\Sigma}_{\bcrv}}

\def\CWg{{\cal W}_{\tsend}}		% bulk causal wedge generalization (CW)
\def\EWg{{\cal W}_{\bcrv}}		% bulk entanglement wedge generalization
\def\CIS{{\cal C}_{\tsend}}		% causal information surface generalization
\def\bcrv{{\cal C}}		% bulk curve
\def\Dhole{{\cal H}}		% bulk hole
\def\extr{{\mathfrak E}}	% extremal surface for a single interval
\def\cong{{\cal N}}		% null congruence
\def\gchi{\chi_\tsend}
\def\CWO{{\blacklozenge}}  % causal wedge for a single observer

\def\ph{\varphi}
\def\rhotp{\rho_{\scriptscriptstyle \subset}}
\def\ttp{t_{\scriptscriptstyle \subset}}
\def\phtp{\ph_{\scriptscriptstyle \subset}}
\def\rhobdy{\rho_{\scriptstyle \infty}}
\def\tbdy{t_{\scriptstyle \infty}}
\def\phbdy{\ph_{\scriptstyle \infty}}
\def\bcpar{\vartheta}
\def\tc{t_{\bcrv}}
\def\rhoc{\rho_{\bcrv}}
\def\phc{\ph_{\bcrv}}

\def\dth{ `}

\title{Covariant Residual Entropy}

\author{Veronika E. Hubeny}

\affiliation{ Centre for Particle Theory \& Department of Mathematical Sciences,\\
Science Laboratories, South Road, Durham DH1 3LE, UK.}\emailAdd{veronika.hubeny@durham.ac.uk}

\abstract{
A recently explored interesting quantity in AdS/CFT, dubbed `residual entropy',  characterizes the amount of collective ignorance associated with either boundary observers restricted to finite time duration, or bulk observers who lack access to a certain spacetime region. 
However, the previously-proposed expression for this quantity involving variation of boundary entanglement entropy (subsequently renamed to `differential entropy') works only in a severely restrictive context.  We explain the key limitations, arguing that in general, differential entropy does not correspond to residual entropy.
Given that the concept of residual entropy as collective ignorance transcends these limitations, we  identify two correspondingly robust, covariantly-defined constructs: a `\GCW' associated with boundary observers and a `\GEW' associated with bulk observers.  These causal sets are well-defined in arbitrary time-dependent asymptotically AdS spacetimes in any number of dimensions.  
We discuss their relation, specifying a criterion for when these two constructs coincide, and prove an inclusion relation for a general case.
We also speculate about the implications for residual entropy.  Curiously, despite each construct admitting a well-defined finite quantity related to the areas of associated bulk surfaces, these quantities are not in one-to-one correspondence with the defining regions of unknown.  This has nontrivial implications about holographic measures of quantum information.
} 

\arxivnumber{}
\begin{document}
\begin{flushright} \small{DCPT-14/25} \end{flushright}

%\begin{flushright} \small{DCTP-13/01} \end{flushright}

\maketitle

\flushbottom
\renewcommand{\thefootnote}{\arabic{footnote}}
% ______________________________________

%~~~~~~~~~~~~~~~~~~~~~~~~~~~~~~~~~~~~~~~~~~~~~~~
\section{Introduction}
\label{s:intro}
%~~~~~~~~~~~~~~~~~~~~~~~~~~~~~~~~~~~~~~~~~~

The last 16 years have evinced the spectacular potential of the AdS/CFT correspondence to elucidate the physics of  strongly coupled field theories on one hand, and of quantum gravity on the other.  Yet, despite the success of numerous diverse applications, we are still grappling with the fundamental underpinnings of the correspondence.  
For instance, we do not yet understand precisely how classical bulk spacetime emerges from the non-gravitational CFT degrees of freedom.  
Spurred by the bold proposal of \cite{Swingle:2009bg,VanRaamsdonk:2009ar,VanRaamsdonk:2010pw}, the notion that the quantum phenomenon of entanglement plays a key role has prompted many explorations over the last few years and gained further impetus with the even more radical notion known as ``ER=EPR"  \cite{Maldacena:2013xja}.

Whatever eventual understanding crystallizes from these explorations, it is already amply clear that there is a deep connection between quantum information theoretic quantities in the field theory and geometrical constructs in the bulk.  In fact, this was already indicated by  the pioneering work of  Ryu \& Takayanagi (RT) \cite{Ryu:2006bv,Ryu:2006ef} who proposed that for static configurations, the entanglement entropy of a given region in field theory is determined by the area of a bulk minimal surface anchored on the boundary of that region.  
However, there are other important measures of entanglement, and other information theoretic quantities, whose bulk duals are not known. 
Likewise, there are very natural bulk constructs, such as the causal wedge and causal holographic information \cite{Hubeny:2012wa}, whose meaning in the field theory is not yet fully understood.\footnote{
While some possibilities for field theoretic interpretation of the causal holographic information \cite{Hubeny:2012wa} were already suggested \cite{Kelly:2013aja} (see also \cite{Freivogel:2013zta}), this programme is still very much in its infancy.
}

When faced with a natural or fundamental quantity on one side of the correspondence, one typically expects that it should have correspondingly natural dual description on the other side.  In attempting to unearth this dual, a basic requirement is for it to be as robust and well-defined as the original quantity.  Conversely, it is of limited utility  to propose quantities which are based on specific coordinate system or pertain only to highly symmetric or otherwise special situations (the fact that we can most easily carry out calculations therein notwithstanding).

For quantities related to bulk geometry, the requirement of general covariance has proved to be an extremely useful guiding principle.  This has played a key role, for instance, in formulating the Hubeny-Rangamani-Takayanagi (HRT) proposal \cite{Hubeny:2007xt} for a covariantly-defined holographic entanglement entropy, valid in a general time-dependent geometry.  The generalization turns out to be an innocuously simple one:  RT's minimal surface `at constant time' gets promoted to a spacetime extremal surface in HRT, which  is also  related to light-sheets used in the covariant entropy bound context  \cite{Bousso:2002ju}, an earlier instance where the requirement of covariance guided the construction.
A more recent example where a robust, natural, and well-defined quantity in arbitrary state inspired the underlying motivation for proposing a correspondingly natural dual (yet to be determined) in the field theory is the aforementioned causal wedge and related constructs of \cite{Hubeny:2012wa}.  Indeed, the requirement of restricting attention to robustly well-defined quantities will provide the basic guiding principle for the present work.

Recently, Balasubramanian et.al.\ \cite{Balasubramanian:2013lsa} proposed an interesting construction for a quantity characterizing the amount of missing information about the full system associated with either boundary observers (on $S^1 \times R$) restricted to finite time duration, or bulk observers (in AdS$_3$) who are causally disconnected from a certain spacetime region in the bulk.\footnote{
Although \cite{Balasubramanian:2013lsa} conflated the two notions, we will see later that they are not only logically distinct, but also generally lead to different constructs.
}  
They originally termed this quantity `residual entropy' (switching the name to `differential entropy' in v2, following its usage in  \cite{Myers:2014jia}) 
and suggested a formula for it in terms of entanglement entropies associated with a family of boundary intervals whose domains of dependence comprise the given time strip.
They were able to show in a restricted context that their expression, which in the continuum limit involves integrating the derivative of entanglement entropy with respect to region size, reproduces the area of the associated bulk `hole', analogously to the Bekenstein-Hawking formula for black hole entropy. 
  This was motivated partly by an earlier work \cite{Balasubramanian:2011wt} which considered the relation between the entanglement across different momentum scales in the CFT to holographic RG flow and made the natural suggestion based on the scale/radius duality that this may be given by the area of spherical surface in the bulk.\footnote{
  Subsequently  \cite{Balasubramanian:2013rqa} analyzed a spherically symmetric hole in AdS$_3$ and confirmed that the gravitational entropy of  resulting `spherical Rindler space' is indeed given by (quarter of) the area of the acceleration horizon which defines the hole.  The work of  \cite{Balasubramanian:2013lsa} generalized this construction to arbitrary curves at constant time AdS$_3$.}
More generally, this bolstered the 
spacetime entanglement conjecture of Bianchi \&  Myers \cite{Bianchi:2012ev} which identified the leading contribution of the {\it bulk} entanglement entropy of a given (arbitrary-shaped) bulk region with the surface area of that region.

Intriguing as the construction of \cite{Balasubramanian:2013lsa} is, its scope is unfortunately severely limited.  As the authors themselves acknowledge, the actual formulation is well-defined only in 3-dimensional bulk, and the explicit derivation pertains only to pure AdS geometry.\footnote{
In a restricted regime, one can generalize to quotients of AdS, such as BTZ, but the formula given in  \cite{Balasubramanian:2013lsa} in terms of entanglement entropies is incorrect unless the boundary time strip has sufficiently short duration.}  Moreover, the interpretation given above does not apply if the boundary time strip or the bulk curve become too `wiggly' -- so even in pure AdS$_3$ and even for short time strips, the construction is not fully robust, in contrast to the abstract concept of residual entropy.  

In a recent endeavor to extend this intriguing construction,
Myers et.al.\ \cite{Myers:2014jia} have generalized  the residual entropy formula of \cite{Balasubramanian:2013lsa} in a number of ways, while retaining the boundary feature of integrating over an expression involving the derivative of entanglement entropy and recovering the gravitational entropy of the bulk region.  In the process they re-named this quantity the `differential entropy' in order to make its definition more suggestive.  The primary extension was
 to translationally-invariant configurations in higher dimensions, by replacing the intervals used in \cite{Balasubramanian:2013lsa} with higher-dimensional strips; but more interestingly, they show explicitly that this construction can be applied to more (though not fully) general asymptotically AdS spacetimes, and even to more general theories of gravity.  
 
 Nevertheless, a number of limitations remain: apart from requiring translational invariance, \cite{Myers:2014jia} only considers static geometries, with the boundary strips lying all at the same time.  
More importantly, the natural interpretation espoused in \cite{Balasubramanian:2013lsa}, of the entropy characterizing the amount of collective ignorance of observers who are causally disconnected from a given bulk spacetime hole, is even further removed from the construction.  On the one hand, one would most naturally associate the former with a causal construction, namely a generalized `causal wedge' pertaining to the given boundary time strip.   Correspondingly, the associated `causal information surface' defined in  \cite{Hubeny:2012wa} would then form the rim of the bulk hole, and its associated quarter-area dubbed `causal holographic information' could then more naturally define the residual entropy.  Indeed, a related proposal has been recently put forward by Kelly \& Wall \cite{Kelly:2013aja}, who suggested the interpretation of causal holographic information as the `one-point-entropy', obtained by maximizing the entanglement entropy over reduced density matrices with one-point functions fixed.
On the other hand, if one blindly applies the formula of \cite{Balasubramanian:2013lsa} (or its generalization discussed in 
\cite{Myers:2014jia}) 
by replacing every occurrence of entanglement entropy with causal holographic information, then as explained in \cite{Myers:2014jia} this does not reproduce the hole's area.\footnote{
  Indeed, as pointed out already in \cite{Hubeny:2012wa} and elaborated in \cite{Freivogel:2013zta,Myers:2014jia}, in higher dimensions the subleading divergences do not cancel, so this quantity would actually be infinite.}
  
However, there is no particular reason to use the formula of \cite{Balasubramanian:2013lsa} in such a construction, since it is not intrinsically well-defined in general.  Indeed, if we use the causal holographic information pertaining to the generalized causal wedge of the full time strip (defined as the bulk region which is causally connected, in both future and past directions, to the boundary time strip), rather than  the union of causal wedges associated with each individual observer, then the corresponding causal holographic information {\it does} give a finite quantity, namely the area of the bulk hole.  This observation partly motivates the present note.  

As remarked above, the original idea of the concept of residual entropy as introduced in \cite{Balasubramanian:2013lsa} transcends all the limitations mentioned above.  We therefore set out to identify a correspondingly robust construct, which is well-defined in a general time-dependent asymptotically AdS spacetime in arbitrary number of dimensions.  The construction is rather simple and the underlying concept is not novel; indeed, our constructs are based purely on causal relations, generalizing the causal wedge discussed in \cite{Hubeny:2012wa} and the entanglement wedge introduced in 
\cite{HHLR14}.  However, it is useful to specify  the construction explicitly, as this serves to analyze its properties and will potentially elucidate the direct meaning on the dual side.
We will adhere to the terminology proposed in both  \cite{Balasubramanian:2013lsa} and \cite{Myers:2014jia}, but distinguish the two concepts:
We will reserve the name {\it differential entropy} for  boundary quantity constructed from derivatives of entanglement entropy for a family of regions, which we will not discuss further.\footnote{
\label{fn:plateaux}
However, it is worth pointing out that it is {\it not} always true that areas of bulk surfaces can be obtained solely from boundary entanglement entropies.
(More precisely, while the area of any closed bulk surface can be obtained by appropriate combination of areas of extremal surfaces anchored on the boundary, the latter need not be the dominant saddles
which compute the holographic entanglement entropy for the said regions.)
} 
 Instead, we will retain the name {\it residual entropy} for the quantity we are interested in: the measure of collective ignorance pertaining to certain inaccessible spacetime region.\footnote{
In fact, both of these terms already exist in literature for describing different quantities in quantum information or condensed matter;
however, as in \cite{Myers:2014jia}, we feel that this context is sufficiently removed to render the danger of confusion due to multiple meanings of the same term smaller than that stemming from multiple terms with the same meaning.}

As mentioned above, there are two natural starting points which specify the realm of unknown, namely the bulk hole and the boundary time strip exterior.
To that end, we propose {\it two} covariantly defined constructs, one associated with each starting point, which naturally implement   the corresponding collective ignorance.  In a tame enough situation, the two constructs coincide, but curiously not in general.   Nevertheless, we will provide a simple criterion for the mismatch  and demonstrate a certain inclusion property that relates the two in full generality.
The main focus of this note will be the behavior of the relevant bulk causal sets and their boundaries.  We will not attempt to define residual entropy within the field theory directly, leaving this for future work.
 
 The plan of this paper is then as follows.  In \sec{s:holeo} we review the `hole-ographic' proposal of \cite{Balasubramanian:2013lsa} and explain its limitations.  In the process we also define the notation and terminology, to pave the way for our proposal.  The main construction of covariantly defined residual entropy is presented in \sec{s:constr}.  
For both starting points (boundary time strip in \sec{s:constrbdy} and bulk hole in \sec{s:constrbulk}), we define co-dimension-zero bulk regions, coining the names {\it \GCW} and {\it \GEW}, respectively.
Since the two constructions may, but need not, coincide, we devote  \sec{s:comp} to explaining their relation.  We conclude with a discussion in \sec{s:discuss}, revisiting possible implications for the residual entropy.  
Since we introduce several new, though related, constructs, which were not previously distinguished, we are compelled to invent new terminology and notation for these; for ease of orientation, we summarize them in \sec{s:notation}.
Furthermore, although we present actual plots to illustrate the constructs, we relegate the derivation of the explicit expressions used for these to \sec{s:geods}, to avoid breaking the flow of the main text.

\paragraph{Note added in v2:} Subsequent to v1 of this paper, the works 
\cite{Czech:2014wka,Headrick:2014eia} appeared which partially address some of the issues mentioned above concerning differential entropy in higher dimensions.  In particular \cite{Czech:2014wka} generalize the construction of \cite{Myers:2014jia} by relaxing translational symmetry (though they still require pure AdS, time-reflection symmetry, and generically the requisite extremal surfaces used are not the ones capturing entanglement entropy for at least some part of the construction), while \cite{Headrick:2014eia} provide a more robust proof of the relation between differential entropy and gravitational entropy in cases with generalized planar symmetry but without time-reflection symmetry.

%~~~~~~~~~~~~~~~~~~~~~~~~~~~~~~~~~~~~~~~~~~~~~~~
\section{Hole-ographic proposal for residual entropy}
\label{s:holeo}
%~~~~~~~~~~~~~~~~~~~~~~~~~~~~~~~~~~~~~~~~~~

Let us start by reviewing the construction of  \cite{Balasubramanian:2013lsa}, phrased in a causal set terminology which will be convenient for the subsequent generalization.
 Broadly-speaking, residual entropy characterizes the amount of collective ignorance shared by a given family of observers, resulting from inaccessibility of a certain spacetime region. 
In particular, \cite{Balasubramanian:2013lsa} consider two types of families of local observers, ones defined in the AdS bulk, and ones within the CFT on the boundary. 
This motivates us to consider two types of causal sets, those pertaining to the full bulk and those just to the boundary.  Denoting the bulk $\bulk$, its boundary $\bdy$, and its closure $\bbulk = \bulk \cup \bdy$,
we use standard notation $J^\pm(p)$ for bulk causal future/past of $p$ (defined as the set of all points in $\bbulk$ which are connected to $p$ by a past/future-directed causal curve), while for the analogous set defined entirely within the boundary, we use the notation $\Jbdy^\pm(p)$.\footnote{
We can also define the interior of these sets, the future/past domain of influence $I^\pm(p)$, correspondingly as the set of points connected to $p$ by timelike curves.
For smooth spacetimes and $p \in \bdy$, the boundary quantities coincide with the restriction of
the corresponding bulk quantities to the boundary, i.e.\ $\Jbdy^\pm(p) = J^\pm(p) \cap \bdy$ and $\Ibdy^\pm(p) = I^\pm(p) \cap \bdy$.
}
Similarly, given some achronal set ${\cal S}$, we use the notation $D^\pm[{\cal S}]$ for the future/past bulk domain of dependence of ${\cal S} \in \bbulk$ (defined as the spacetime region from which any inextendible causal curve must intersect ${\cal S}$)
 and correspondingly $\Dbdy^\pm[{\cal S}]$ for the boundary domain of dependence of ${\cal S} \in \bdy$.  We also define the full causal development
 $D=D^+ \cup D^-$  and similarly for $\Dbdy$.

%-------------------------------
\subsection{Construction}
\label{s:diffentconstr}
%-----------------------

Following \cite{Balasubramanian:2013lsa}, we first restrict attention to pure global AdS$_3$ (WLOG we set AdS radius $=1$), 
\begin{equation}
ds^2 = - (r^2+1) \, dt^2 + \frac{dr^2}{r^2+1} + r^2 \, d \ph^2
\label{globAdSmetr}
\end{equation}	
 so the boundary CFT lives on the Einstein Static Universe $S^1 \times R$, with
 $ ds_{bdy}^2 = - dt^2 + d\ph^2$.
In the bulk, one can specify the region of inaccessibility by a simple
closed spacelike curve $\bcrv$ at $t=0$; this defines  a `hole' $\Dhole$ in the interior of AdS, corresponding to the bulk domain of dependence  of a spacelike co-dimension-1 region enclosed by $\bcrv$.
However, regardless of the size and spatial position of $\Dhole$, any bulk observer in AdS$_3$ will come into causal contact with $\Dhole$ after at most the time $t_{max}  = \pi$; so to avoid this, the observers must terminate on the AdS boundary earlier.
To  construct a family of observers who are causally disconnected from $\Dhole$, we can then proceed as follows.  Parameterizing $\bcrv$ by a parameter $\bcpar \in [0, 2\pi)$,  at each $\bcpar$ consider the two outgoing (future and past directed) null geodesics normal to $\bcrv$.
The endpoints of these null geodesics will lie on the AdS boundary at 
$p_\pm(\bcpar) = (t=\pm\tbdy(\bcpar), r=\infty, \ph = \phbdy(\bcpar))$
where the time $\tbdy(\bcpar)$ and angle $\phbdy(\bcpar)$ can be easily obtained from $\bcrv$. 

If $\phbdy(\bcpar) \in [0, 2\pi)$ is a monotonically increasing function of $\bcpar$, then any observer $O_\bcpar$ following a timelike trajectory which starts at $p_-(\bcpar)$ and ends at $p_+(\bcpar)$ will be causally disconnected from $\Dhole$;
in fact, that observer's Rindler horizon\footnote{
Any such bulk observer whose worldline ends at $p_\pm$ on the AdS boundary is eternally accelerating and has an associated Rindler horizon.  As discussed below, this horizon corresponds to the boundary of that observer's causal wedge, and in pure AdS$_3$, its spatial section at $t=0$ is a bulk extremal surface (i.e.\ a spacelike geodesic).
} is tangent to $\bcrv$ at $\bcpar$.  Since each individual observer is causally disconnected from $\Dhole$, so is the entire family of such observers, parameterized by $\bcpar$.  The idea of residual entropy phrased in \cite{Balasubramanian:2013rqa,Balasubramanian:2013lsa} as collective ignorance of bulk observers pertains to this family.
On the other hand, if $\phbdy(\bcpar)$ is not monotonically increasing, then at least some observers will come into causal contact with $\Dhole$ -- in other words, they can influence or be influenced by the physics inside the hole.  We will postpone the discussion of these more subtle cases till after establishing the basics.

Let us for the moment consider the former case, of all bulk observers $O_\bcpar$  being causally disconnected from $\Dhole$, as is implicitly assumed in the motivation of the residual entropy concept.  Since $\phbdy(\bcpar)$ is monotonically increasing,
$\tbdy$ is a single-valued function of $\bcpar$ (or equivalently of $\phbdy$).  The set of future endpoints $\tsend^+ = \{ p_+ \}$  and the set of past endpoints  $\tsend^- = \{ p_- \}$ then form smooth spacelike curves on the AdS boundary, which bound a boundary spacetime region $\ts = \{ (t,\ph) : -\tbdy(\ph) \le t \le \tbdy(\ph) \}$.  This region, which \cite{Balasubramanian:2013lsa} refer to as the `time strip', can be defined more generally as 
\begin{equation}
\ts = \Jbdy^-[\tsend^+] \cap \Jbdy^+[\tsend^-]
\label{tsdef}
\end{equation}	
with $\tsend^\pm$ forming its future/past boundaries.

\begin{figure}
\begin{center}
\hspace{1.5cm}
\includegraphics[width=.32\textwidth]{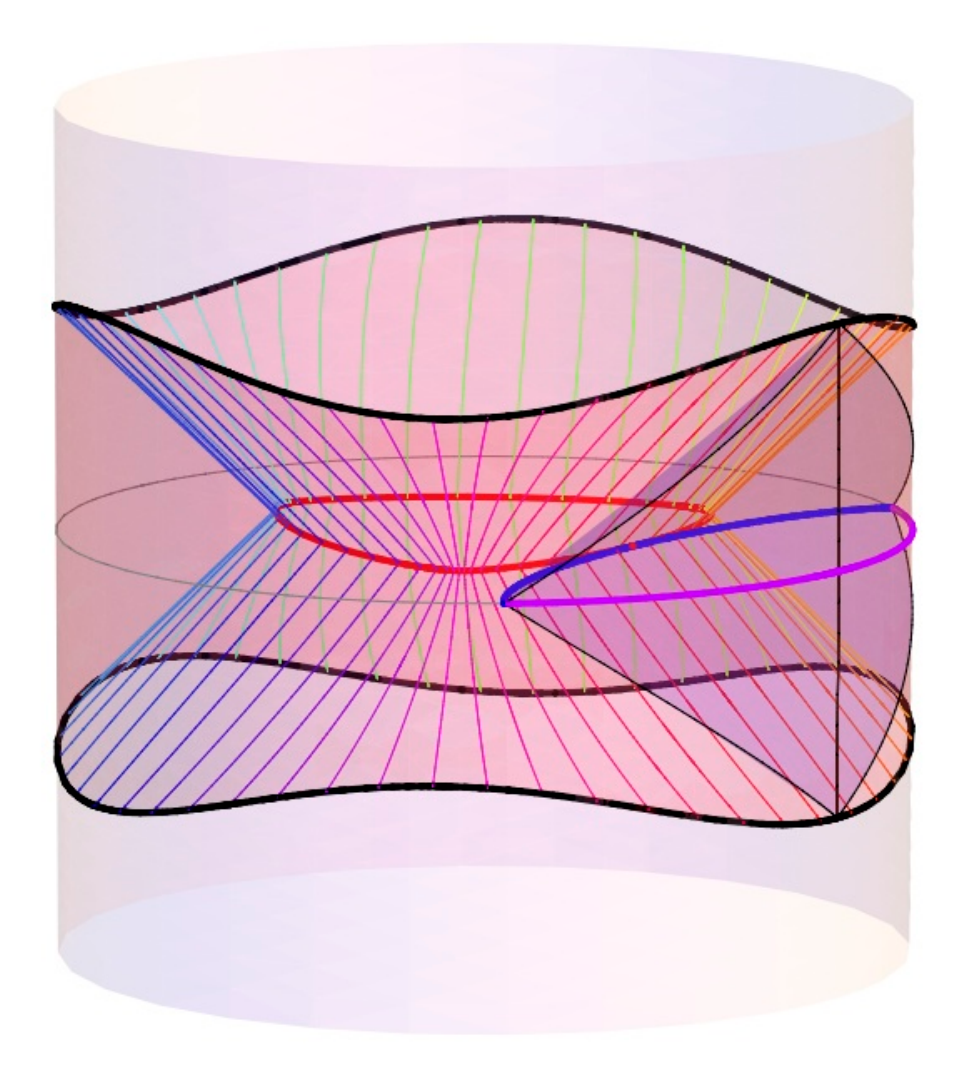}
\hspace{1cm}
\includegraphics[width=.45\textwidth]{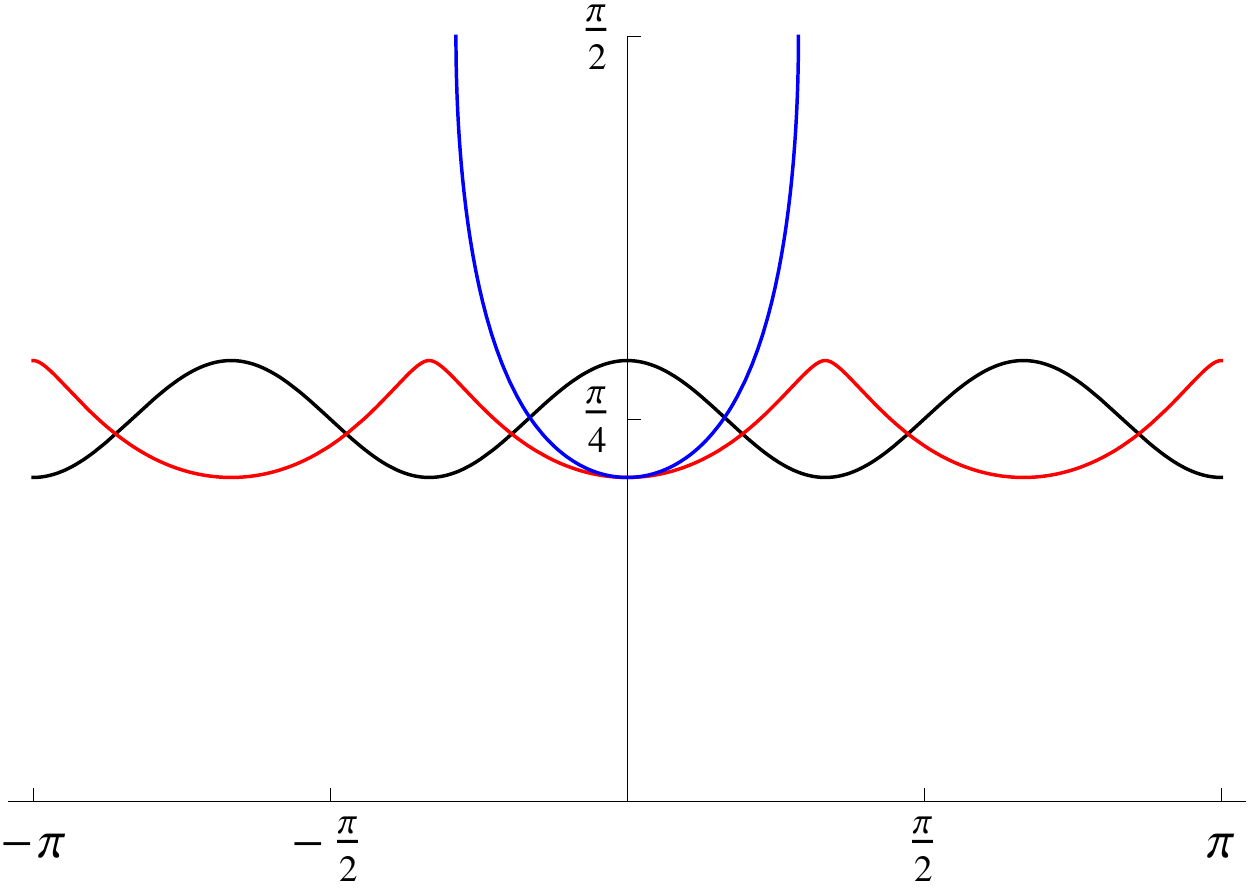}
\hspace{0.5cm}
\begin{picture}(0,0)
\setlength{\unitlength}{1cm}
\put(-0.6,0.5){$\ph$}
\end{picture}
\caption{
Relation between bulk curve (bounding the inaccessible hole) and boundary time strip in residual entropy construction in AdS$_3$ for simplest type of scenario.  
{\bf Left:} AdS$_3$ diagram, with bulk hole rim $\bcrv$ [thick red curve], its null normals [lines color-coded by $\phbdy$] whose endpoints determine the extent the boundary time-strip $\ts$ [shaded red] bounded by $\tsend^\pm$ [thick black curves].  We also indicate boundary $t=0$ slice [thin gray curve], one boundary observer at $\bcpar=0$ [vertical black line] and his causal wedge; the latter is obtained from the construction of \cite{Hubeny:2012wa} applied to the interval $I_0$ [thick purple curve] defining the boundary observer's causal domain.  The causal information surface [thick blue curve] coincides with the extremal surface $\extr_0$ whose area defines the entanglement entropy $S(I_0)$.
The axis are $(\rho \, \cos \ph, \rho \, \sin \ph, t)$ where $\rho = \tan^{-1} r$, so that radial null geodesics are inclined at 45 degrees.
{\bf Right:} The corresponding functions of $\ph$ describing the temporal profile of the time strip boundary $\tsend$ [black], the radial profile of the curve $\bcrv$ [red], and the radial profile of the extremal surface $\extr_0$ [blue] which ends on $\partial I_0$.  
Note that $\extr_0$ is tangent to $\bcrv$ at $\bcpar=0$ and strictly outside $\Dhole$ everywhere else.
}
\label{f:TS}
\end{center}
\end{figure}

For example, in the simplest case of spherically symmetric $\bcrv$ lying not only at constant $t=0$ but also at constant $r=r_0$ (which was the case studied in \cite{Balasubramanian:2013rqa}),
 the null normals are simply radial geodesics so that $\phbdy(\bcpar) = \bcpar$, and they all reach the boundary at the same time, 
$\tbdy(\bcpar) = \pi/2 - \tan^{-1} r_0$, so in this case the time strip is of constant duration everywhere.  For a non-spherically-symmetric $\bcrv$, the null normals will reach the boundary at different times.  This is illustrated in the spacetime plot in the left panel of \fig{f:TS} which indicates this construction in a generic example within this restricted class: $\bcrv$ (thick red curve) lies at $t=0$ and is such that $\tsend^\pm$ (thick black curves) formed by the endpoints of the null normals (thin lines color-coded by $\phbdy$) give a smooth single-valued function $\tbdy(\phbdy)$.  This function and the radial profile of $\bcrv$ are plotted in the right panel of \fig{f:TS} (with corresponding color-coding).  The time strip $\ts$ is the shaded region lying between $\tsend^+$ and $\tsend^-$.  The remaining parts of the figure will be discussed below.

% Figure --------
% \fig{f:TS} --

The restriction of boundary spacetime to the time strip $\ts$ then motivates a related boundary quantity, now defined in terms of boundary observers:  By interpreting the time strip $\ts$ as the time interval during which boundary observers can make measurements, the (boundary) residual entropy characterizes the collective ignorance resulting from this temporal restriction.  As in the case of the bulk family of observers, here too the specific  family of  boundary observers is not uniquely defined.  However, each observer's starting and ending point is implicitly assumed to be given by $p_\pm(\bcpar)$ and the observers again to be parameterized by $\bcpar$.  
Note that since we took $\bcrv$ to lie at constant time $t=0$, by flip symmetry $t \to -t$,  the angular ($\ph$) value of $p_+$ matches that of $p_-$ for each $\bcpar$, so we could  imagine that each observer is a static one, sitting at a specified angle $\ph$. (Such an observer for $\bcpar=0$ is indicated by the vertical black line on the spacetime plot of \fig{f:TS}.)

The authors of \cite{Balasubramanian:2013lsa} went on to convert this temporally-induced collective ignorance into a spatially-induced one, by rephrasing this in terms of reduced density matrices for a family of overlapping spatial intervals at $t=0$. 
In particular, they defined the residual entropy as the maximal entropy of the density matrix for the full system consistent with all observations made by the specified family of local boundary observers. 
Each observer is allowed to make observations only within a restricted region, defined either by the endpoints $p_\pm(\bcpar)$ or equivalently by an interval $I_\bcpar$ on the boundary at $t=0$, whose domain of dependence implements the temporal restriction,
\begin{equation}
\Dbdy[I_\bcpar] = \Jbdy^-(p_+(\bcpar)) \cap  \Jbdy^+(p_-(\bcpar))
\label{tsdefp}
\end{equation}	
For example, \fig{f:TS} shows the  $\bcpar=0$ interval $I_0$  by the thick purple curve (which is a subset of the boundary $t=0$ slice, indicated by the thin gray curve).  The corresponding domain of dependence $D[I_0]$ is also indicated.

Hence we have now converted the bulk curve $\bcrv$ at $t=0$ to a family of boundary intervals $I_\bcpar$ at $t=0$, to each of which one can associate the entanglement entropy $S(I_\bcpar)$, defined as the Von Neumann entropy of the reduced density matrix $\rho_\bcpar$, obtained by tracing the full density matrix over the degrees of freedom outside $I_\bcpar$ as usual.
Using strong subadditivity to motivate the construction for a discrete family of boundary observers, \cite{Balasubramanian:2013lsa} then define the residual entropy for a collection of intervals $I_k$ as 
\begin{equation}
E = \sum_k \left[ S(I_k) - S(I_k \cap I_{k+1}) \right] \ .
\label{diffentdiscr}
\end{equation}	
Even though each $S(I_k)$ is UV-divergent, in the full expression \req{diffentdiscr} these divergences cancel, yielding a finite quantity.  (This is because each endpoint responsible for the divergence enters once with a plus sign and once with a minus sign.)
In the continuum limit where the intervals vary continuously in  position and size, $E$ can be expressed as an integral, with the integrand involving the derivative of the entanglement entropy with respect to the region size,
\begin{equation}
E = \frac{1}{2} \, \int_0^{2\pi} d\ph \, \frac{d S(\alpha)}{d \alpha} \mid_{\alpha(\ph)}
\label{diffentcont}
\end{equation}	
where $\alpha(\ph)$ corresponds to the size of the interval centered at $\ph$.

Having described an interesting boundary quantity, let us return to its bulk description.  The bulk construct which encapsulates each entanglement entropy $S(I_\bcpar)$ is the quarter-area of an extremal surface $\extr_\bcpar$ (which in this 3-dimensional case is a spacelike geodesic) anchored on the entangling surface $\partial I_\bcpar$,
i.e.\
\begin{equation}
S(I_\bcpar) = \frac{{\rm Area}(\extr_\bcpar)}{4 \, G_N}
\label{HRTrel}
\end{equation}	
Such a surface for $\bcpar=0$ is indicated by the blue curve in both panels of \fig{f:TS}.  As explained in  \cite{Hubeny:2012wa}, in  pure AdS$_3$, the extremal surface  
 in fact coincides with the rim of the corresponding causal wedge (also indicated in \fig{f:TS}), dubbed  `causal information surface' in  \cite{Hubeny:2012wa}.  That means that it  coincides with the given bulk observer's Rindler horizon at $t=0$.
 Note that as mentioned above, it is by construction tangent to the bulk curve $\bcrv$.

Using \req{HRTrel}, \cite{Balasubramanian:2013lsa} were able to show that the full integral \req{diffentcont} reproduces the quarter-area of the original bulk curve $\bcrv$,
\begin{equation}
E = \frac{{\rm Area}(\bcrv)}{4 \, G_N}
\label{diffentarea}
\end{equation}	
This was further explained and generalized in \cite{Myers:2014jia}, who called $E$ the differential entropy.
Geometrically, one may think of this as piecing together the bulk curve $\bcrv$ from incremental segments of the extremal surfaces $\extr_\bcpar$ at their tangent point, with the remaining parts' areas canceling out by subtracting the nearby extremal surfaces for the intersecting intervals.  This cancellation is actually non-trivial, and generically involves subtracting off a total derivative from the integrand in \req{diffentcont}.
Remarkably, this relation remains valid even in the regime of non-monotonic $\phbdy(\bcpar)$, where the time strip interpretation no longer holds.  We will return to this point below. 

Note that despite \cite{Balasubramanian:2013lsa}'s  original motivation, the final construction yielding \req{diffentarea} has now been rephrased in a way which dispenses with causal constructs entirely: instead of talking about the boundary time strip, one simply considers a family of extremal surfaces $\extr_\bcpar$ which are tangent to $\bcrv$.  Their endpoints define the family of boundary intervals $I_\bcpar$ to which the construction \req{diffentdiscr} (or its continuum limit \req{diffentcont}) applies.

%-------------------------------
\subsection{Limitations}
\label{s:limitations}
%-----------------------

Having reviewed the basic construction of  \cite{Balasubramanian:2013lsa}, let us now turn to its main limitations.  We first list them, and then discuss each in turn.  Although some of these points were already acknowledged in \cite{Balasubramanian:2013lsa}, and some were further explained in  \cite{Myers:2014jia}, here we'll nevertheless include them for completeness and later convenience in explicating our proposed constructions in \sec{s:constr}.

\noindent
The construction of \cite{Balasubramanian:2013lsa} is formulated:
\begin{enumerate}
\item
only in 2+1 dimensions,
\item
only for pure AdS geometry,
\item
only for bulk curve $\bcrv$ which lies at constant time, or equivalently for  boundary time strip $\ts$ which is time-flip-symmetric,
\item
only for `tame enough' situations, e.g.\ bulk curve $\bcrv$ (or the time strip boundary $\tsend$) which is not too wiggly (specified more precisely below).
\end{enumerate}
The first two points were partially addressed by  \cite{Myers:2014jia}, but with further limiting restrictions; for example limitation 3 pertained in that case as well, apart from the extra requirement of translational invariance and therefore planar AdS.  
Limitation 3 has been surmounted very recently in \cite{Headrick:2014eia} (with further adjustments), while limitation 4 remains for {\it all} developments hitherto, as does the fact that the extremal surfaces used in the construction are not always the ones relevant for entanglement entropy, so that the differential entropy expression formulated in terms of entanglement entropy is not universally valid.
More importantly, however, even with all the generalizations presented to date, the proposed differential entropy does {\it not} generically give a residual entropy, unless all of the above are satisfied.

%----------------
\paragraph{1. Dimensionality:} 
First of all, note that a bulk hole $\Dhole$ only makes sense as a co-dimension-zero spacetime region, so that the defining bulk `curve' $\bcrv$ must generically be a bulk co-dimension-two surface.  
For AdS$_{d+1}$, $\bcrv$ is then a $d-1$ dimensional spacelike surface, as is any extremal surface $\extr$ relevant for  boundary entanglement entropy.  
Generalizing \req{diffentdiscr} would then involve replacing the 1-dimensional intervals $I_\bcpar$ with $d-1$ dimensional regions which cover the boundary $t=0$  slice.

The most obvious reason why 3-dimensional spacetime (i.e.\ $d=2$) was essential for the formulation of \cite{Balasubramanian:2013lsa} is that \req{diffentdiscr} requires an ordering of the boundary intervals in order to define their intersections, or in the continuum limit \req{diffentcont} to define the requisite derivative.  In higher dimensions, there is no natural ordering of compact regions, nor a natural direction for a derivative.  Likewise, given a point ${\vec \bcpar}$ on some bulk surface $\bcrv$, there is no longer a uniquely-defined tangent extremal surface $\extr_{{\vec \bcpar}}$; rather there are infinitely many tangent extremal surfaces, specified by their extrinsic curvature at the tangent point.

To circumvent the former obstacle,  \cite{Myers:2014jia} considered infinite strips to restore this ordering, but that came with the price of requiring translational symmetry along the strips,\footnote{
It may be possible to relax the requirement translational invariance as well by considering non-uniform strips \cite{MSVH}, but even so, it loses some degree of naturalness.
} which simultaneously circumvents the latter obstacle.
  With this extension, \cite{Myers:2014jia} could show that the relation between differential entropy and bulk area carries through, i.e.\ that the family of extremal surfaces which are tangent to the given surface $\bcrv$ have entanglement entropies which reproduce the area of $\bcrv$ via a generalization of \req{diffentdiscr}, essentially for the same geometrical reason as indicated at the end of \sec{s:diffentconstr}.

However, such tangent extremal surfaces do {\it not} implement the correct (causally-determined) time strip corresponding to $\bcrv$.  In particular, the boundary domains of dependence for the regions defined by these tangent surfaces  extend in time only commensurately with the region width.  But while causally disconnected from the bulk `hole', it is not the maximal such region: null normals from $\bcrv$ reach the boundary an ${\cal O}(1)$ distance outside the putative time strip.\footnote{
More specifically, the discrepancy is given by a factor of
$ \frac{2 \, (d-1)}{\sqrt{\pi}} \, 
\Gamma \left( \frac{2d-1}{2d-2} \right) / \Gamma \left( \frac{d}{2d-2} \right) $; see e.g.\ Fig.5 of \cite{Hubeny:2012ry} for the plot of the radial profile of such a surface for various $d$, which demonstrates the fact that co-dimension-two surfaces anchored on a strip of fixed width reach deeper for larger $d$.
}

Conversely, suppose one started with the boundary time strip $\ts$ defined by $\bcrv$ as given in \req{tsdefp}.  
Then first of all, a set of boundary local static observers require round ball boundary regions, not infinite strips, in order to recover the correct domain of dependence associated with their endpoints; and second of all, the corresponding bulk extremal surfaces anchored on these round regions do not reach all the way to $\bcrv$.  Instead, in pure AdS$_{d+1}$, such extremal surfaces coincide with the causal information surface, consistently with the causal wedge interpretation.  

Given this, one might be tempted to try constructing some formula for residual entropy based on these local observers' causal information surfaces.  However, in addition to being faced with the ordering ambiguity, here we would also encounter the problem of divergences, explained in \cite{Myers:2014jia} (see also  \cite{Hubeny:2012wa,Freivogel:2013zta}): the causal holographic information $\chi$ defined as the quarter-area of the causal information surface \cite{Hubeny:2012wa} has subleading UV divergences which depend not only on the local geometry of the entangling surface but also on more delocalized information pertaining to the  full domain of dependence of the given region; these divergences then generically do not  cancel in expressions analogous to \req{diffentdiscr}.
 
%----------------
\paragraph{2. Geometry:} The requirement of pure AdS geometry in \cite{Balasubramanian:2013lsa} was used mainly for convenience in implementing the explicit calculations, since as shown in \cite{Myers:2014jia}, the differential entropy expression \req{diffentcont} satisfies  \req{diffentarea} more generally as well.  In fact, even here there is a caveat to worry about, related to the remark in footnote \ref{fn:plateaux}:
There are certain situations, such as large regions (long time strips) in a black hole background, where the entanglement entropy is not given by the desired extremal surface in the geometrical construction of \cite{Balasubramanian:2013lsa,Myers:2014jia}, but rather by a pair of surfaces, one homologous to the complement region and one wrapping the black hole.  This holds quite robustly \cite{Hubeny:2013gta}, and implies that there is a sudden transition when the expression \req{diffentcont} ceases to reproduce \req{diffentarea}.
For example, for BTZ black hole of size $1/2$ AdS radius, this happens when the size of the system is larger than $\sim 0.79$ of the total volume, i.e.\ the time strip has total duration longer than $\sim 2.5$ in AdS units.  (This is a significant restriction, since unlike the case in pure AdS, here for a bulk region surrounding the black hole sufficiently near to the horizon, the corresponding time strip could be arbitrarily long.)

However, generalizations to other backgrounds aside, the main motivation of \cite{Balasubramanian:2013lsa} for working in pure AdS$_3$ was the equivalence between extremal surfaces and causal information surfaces.  As explained in  \cite{Hubeny:2012wa}, this coincidence fails (for any boundary region) in more general asymptotically AdS$_3$ spacetimes which are not locally AdS (as well as non-round regions in higher-dimensional AdS, as discussed above).  Hence for general geometries, we encounter the same problem as for limitation 1, namely the direct generalization of the differential entropy defined by \req{diffentdiscr} or  \req{diffentcont} does not recover a residual entropy.

Even more disconcertingly, for general time-dependent geometries, we can no longer meaningfully specify a ``$t=0$" bulk slice on which to define $\bcrv$, since unlike the static case, the coordinate $t$ here does not have a natural geometric meaning.  
Relatedly, a generic boundary time strip $\ts$ corresponding to some  bulk curve $\bcrv$ will not be time-flip symmetric, which leads to an observer selection problem discussed in the next item.
Moreover, even if we ignore the time strip and just use the final construction involving tangent geodesics to the curve $\bcrv$ in 3 dimensions, the corresponding endpoints of $\extr_\bcpar$ will no longer lie on the same boundary time-slice, which means that the differential entropy formula cannot involve entanglement entropies at a given time, rendering \req{diffentdiscr} and \req{diffentcont} ill-defined.

%----------------
\paragraph{3. Time-symmetry:} 
To implement a time-flip-symmetric construction, one needs not only a static spacetime but also a bulk curve $\bcrv$ lying at constant time.  This naturally allows for considering static observers on the boundary.  
More importantly, it also guarantees that all corresponding intervals $I_\bcpar$ lie at the same boundary slice $t=0$, so in particular one can meaningfully take their intersections in \req{diffentdiscr}.  However this will no longer hold once time symmetry is broken.

To illustrate the point, it suffices to consider the case of planar 
AdS$_3$.  Even in this simple case of static spacetime, time-flip symmetry can be broken both by a non-static family of boundary observers, and by static observers on non-time-flip-symmetric time strip (-- for uniform strip and inertial observers these are in fact related by a boost).  Let us first focus on the reach of the causal wedge associated with a given observer, and  then compare the reach for various observers in a given time strip.
Take Poincare-AdS, written in the standard Poincare coordinates 
\begin{equation}
ds^2  = \frac{-dt^2+dx^2+dz^2}{z^2}
\label{}
\end{equation}	
and a boundary observer whose worldline has endpoints 
\begin{equation}
p_\pm = ( \, t = \pm t_0 \,  ,  \, x = \pm x_0 \, ,  \, z=0 \, )
\label{genendpoints}
\end{equation}	
with $0\le x_0<t_0$.
Such an observer has an associated causal development enclosed by future-directed null rays from $p_-$ and past-directed null rays from $p_+$
 (these are given by  the lines $t=\pm t_0\pm(x-x_0)$),
 which intersect at 
$r_\pm =  ( \, t = \pm x_0 \,  ,  \, x = \pm t_0 \, ,  \, z=0 \, )$.  
The intersection points $r_\pm$ specify the endpoints of a spacelike interval $I_0$ whose domain of dependence gives the requisite causal development, 
$\Dbdy[I_0] = \Jbdy^+(p_-) \cap \Jbdy^-(p_+)$.  Correspondingly, we can define the associated causal wedge in the bulk by the intersection of bulk light cones from $p_\pm$, i.e.\
$J^+(p_-) \cap J^-(p_+)$.
Its future/past boundaries are then given by 
$(t \pm t_0)^2 = (x \pm x_0)^2 + z^2$, 
which means that the causal information surface, defined by the intersection of the two light cones, has the radial and temporal profile respectively given by
\begin{equation}
z(x) = \sqrt{\left( t_0^2-x_0^2 \right) \, \left(  1 - \frac{x^2}{t_0^2} \right) }
\ , \qquad 
t(x) = \frac{x_0}{t_0} \, x  \ .
\label{geodradprofboost}
\end{equation}	
This gives a spacelike curve anchored on $\partial I_0 = \{ r_\pm \}$, 
which is indeed precisely the bulk spacelike geodesic $\extr_0$
joining $r_+$ and $r_-$.

\begin{figure}
\begin{center}
\includegraphics[width=.9\textwidth]{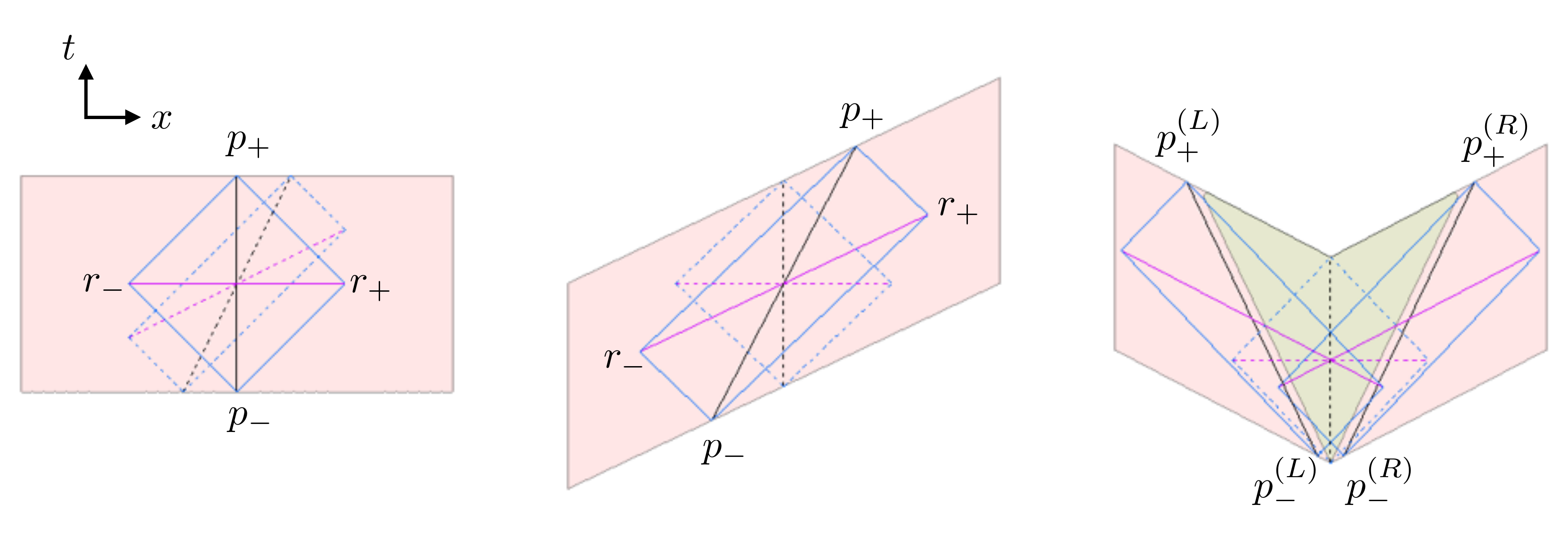}
\caption{Various boundary time strips [shaded region] and corresponding `optimal' observers [solid, endpoints labeled by $p_\pm$] versus non-optimal observers [dashed].  For each observer [black line], we also show the corresponding causal development [bounded by the blue diagonal lines] and interval $I_0$ [purple line, for optimal observers  labeled by endpoints $r_\pm$]. 
{\bf Left:} For time-flip-symmetric uniform time strip, static observers are optimal.
{\bf Middle:} For a boosted time strip, correspondingly boosted observers are optimal.
{\bf Right:}
For more complicated time strip, generically the intervals $I_\bcpar$ do not lie on the same spatial slice of the boundary.  Moreover, there can be regions [shaded green] which are not traversed by longest-lived observers.
}
\label{f:bdyTSobs}
\end{center}
\end{figure}

The main result of this simple calculation, which  follows immediately from \req{geodradprofboost}, is that the deepest reach of $\extr_0$ into the bulk is 
$z_0 = \sqrt{t_0^2-x_0^2}$.  Note that for fixed $t_0$, this is clearly maximized when $x_0=0$, corresponding to a static observer.  
This means that for constant-duration time strip 
\begin{equation}
\ts = \{  (t,x) \mid -t_0 \le t \le t_0 \} \ ,
\label{DstraightPoinc}
\end{equation}	
a family of static observers considered in \cite{Balasubramanian:2013lsa} would have the deepest associated extremal surfaces $\extr$, with corresponding $\bcrv$ at $z_0 = t_0$.    Any set of boosted observers \req{genendpoints} with nonzero $x_0$ would have shallower causal wedges, and equivalently shallower extremal surfaces.  Moreover, only for static observers do all boundary intervals lie at the same time slice, namely $t=0$.
This is demonstrated in the left panel of \fig{f:bdyTSobs}, where the solid curves correspond to the optimal (static) observer with endpoints $p_\pm$, while dashed ones correspond to non-optimal (boosted) observer.  Note that we indicated  the observer's worldline by a straight line (describing an inertial observer) and likewise we considered the interval $I_0$ as straight (i.e.\ lying at constant time in some boosted frame), but neither of these simplifications is essential: only the respective endpoints $p_\pm$ and $r_\pm$ matter.

On the other hand, suppose we instead have a boosted time strip  
\begin{equation}
\ts_\alpha = \{  (t,x) \mid -{\tilde t}_0 +\alpha \, x \le t \le {\tilde t}_0 +\alpha \, x\} \ ,
\label{DtiltPoinc}
\end{equation}	
with $|\alpha| < 1$, such as shown in the middle panel of \fig{f:bdyTSobs}.
Static observers with $x_0=0$ (dashed lines) would then have associated extremal surfaces which again reach to $z_0 = {\tilde t}_0$, but these are no longer the optimal observers:  a suitably boosted family of observers would yield deeper extremal surfaces.  In particular, an observer whose trajectory is described by $t= \beta \, x$ would intersect the strip boundaries at $t_0 = \beta \, x_0 = {\tilde t}_0 +\alpha \, x_0$, so the associated reach $z_0$ is maximized when $x_0 = \frac{\alpha}{1-\alpha^2} \, {\tilde t}_0  = \alpha \, t_0$.  In other words, this corresponds to a boosted observer (solid lines) with boost $\beta = 1/\alpha$,  the same boost which obtains the tilted strip \req{DtiltPoinc} from a horizontal one \req{DstraightPoinc}.  Such boosted observers' causal wedges reach to $z_0 = \frac{{\tilde t}_0}{\sqrt{1-\alpha^2}}\ge {\tilde t}_0$.  Moreover, it is precisely these observers for whom the corresponding intervals $I_\bcpar$ lie on the same spacelike slice, in this case given by the equation $t=\alpha \, x$.  This example illustrates the lesson that to implement the prescription of \cite{Balasubramanian:2013lsa}, we cannot restrict attention to static observers.

% Figure 
% \fig{f:bdyTSobs}

However, the above case of a tilted strip is deceptively simple, because it is just a boosted version of the time-flip symmetric case.  For generic $\tsend^\pm$, this will no longer be the case.  In principle, to determine the optimal set of observers for a generic time strip, we should choose a family maximizing the reach of the corresponding extremal surfaces. Based on the above examples, we would expect this family to be one maximizing each observer's proper time within the time strip (since both the reach and the proper time are boost-invariant quantities, with the reach $z_0$ given by half of the proper time of an inertial observer ending on $p_\pm$).  However, one might justifiably worry whether this is guaranteed give a proper congruence of observers, i.e.\ whether  the time strip will be foliated by such a family.  It is not difficult to argue that distinct observer worldlines cannot cross because that would contradict maximizing their proper time; however it is not clear that all points within the time strip would lie on some observer's worldline.  

A simple counter-example to foliating observers is illustrated in the right panel of  \fig{f:bdyTSobs}, which is just a reflected version of half of the boosted time strip of the middle panel.  
Hence optimal boosted observers with $p_-^{(R)}$ in the right half coincide with the optimal observers indicated in the middle panel.  Similarly, by reflection symmetry the optimal boosted observers with $p_-^{(L)}$ in the left half are the reflected versions of the right observers, i.e.\ have opposite boost.  This however leaves a region in the middle, indicated by the green-shaded region in \fig{f:bdyTSobs}, through which there are no longest-lived observer worldlines.  E.g.\ a static observer (dashed lines) lives a shorter time than slightly boosted observer originating from the same starting point.
Moreover, as is also evident from this example, even for the family of longest-lived observers, the endpoints $r_\pm$ of the corresponding intervals cannot lie on the same time slice, since $r_+^{(L)}$ (right endpoint of the left purple line) lies in the past of the slice (containing the right purple line) generated by the endpoints $r_\pm^{(R)}$.  This renders the differential entropy expression \req{diffentdiscr} ill-defined.

The above explicit example aside, it is simple to see why  the set of optimal boundary intervals $I_\bcpar$ do not generically overlap.  If we start with a smooth closed bulk curve $\bcrv$ which does not lie at constant time, and at each $\bcpar$ construct a tangent geodesic, we get {\it two} curves on the boundary, one generated by one set of endpoints $r_+(\bcpar)$ and one by the other set $r_-(\bcpar)$.  Both curves are piecewise\footnote{
As we discuss in point 4 below, there can be isolated cusps where the curve doubles back.
} smooth and spacelike, and each covers the entire space on the boundary, but the temporal position of $r_+(\ph)$ and $r_-(\ph)$ at a given angle $\ph$ is determined by the tangent to $\bcrv$ at different points $\bcpar$, which are independent of each other.  As mentioned above, this makes the differential entropy prescription as it stands inapplicable: in the discrete formulation \req{diffentdiscr} the intersections $I_k \cap I_{k+1}$  generically are not intervals so that the UV divergences in components of $E$ cannot cancel, while in the continuous formulation \req{diffentcont} we would presumably need to consider derivative on entanglement entropy with respect to time as well as size.

Throughout this discussion we have been dealing with the simplest case of pure AdS$_3$, which is static and we don't face the difficulties mentioned in points 1 and 2.  It should be clear that for more general spacetimes, the problems mentioned here would be severely compounded.  For example,  
already in stationary spacetime such as rotating BTZ (which is in fact still locally AdS$_3$), different geodesics $\extr$ anchored on the {\it same} boundary time slice $t=0$ do {\it not} all lie on the same bulk spacelike slice, as discussed in \cite{Hubeny:2007xt}; so conversely even for $\bcrv$ at constant bulk time $t=0$, the set of tangent geodesics will not have both endpoints at equal times.

%----------------
\paragraph{4. Shape restrictions:} 
We now come to  the most interesting limitation, but one which has hitherto received the least attention.
Recall that when reviewing the construction of \cite{Balasubramanian:2013lsa} in \sec{s:diffentconstr}, we focused on the case of monotonic $\phbdy(\bcpar)$, as exemplified in \fig{f:TS}.  This allowed for a certain `reversibility of construction', tacitly assumed in \cite{Balasubramanian:2013lsa}:  the time strip $\ts$ generated from $\bcrv$ has associated family of boundary intervals $I_\bcpar$ with corresponding extremal surfaces whose envelope {\it reproduces} $\bcrv$.  The underlying reason, which we revisit more extensively in \sec{s:comp}, is that  in this case the null congruences which determine $\tsend^\pm$ from $\bcrv$ are precisely the same as the null congruences which determine $\bcrv$ form  $\tsend^\pm$.  
However, this reversibility can in fact fail under rather mild conditions, namely when the null generators start to intersect, which happens generically when the congruence becomes too long or the defining curve too wiggly.
Since there are two natural starting points for constructing residual entropy, namely the bulk hole specified by $\bcrv$ and the boundary time strip defined by $\tsend^\pm$, there will be two separate subtleties when either of these crosses a certain threshold.

\begin{figure}
\begin{center}
\hspace{1.5cm}
\includegraphics[width=.3\textwidth]{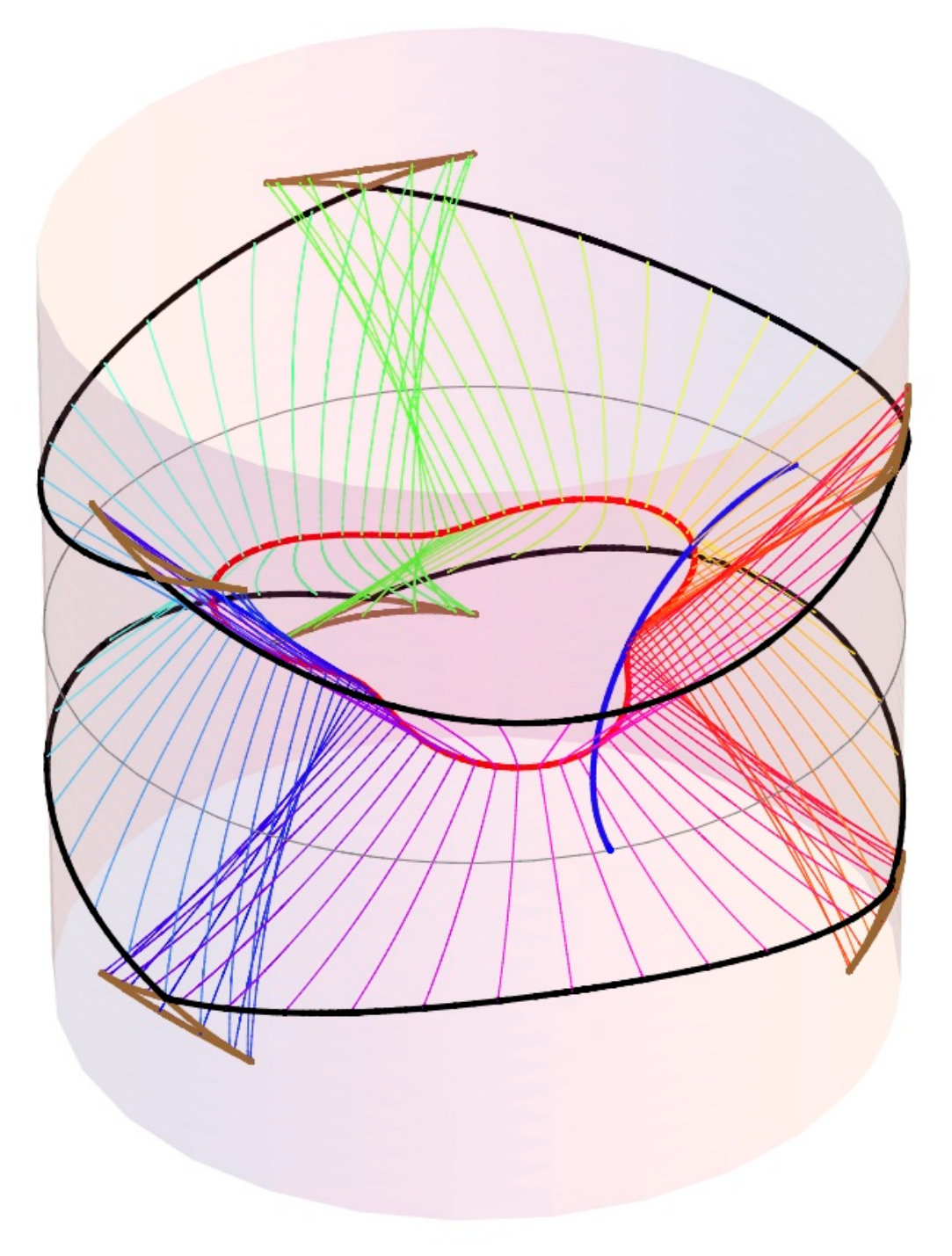}
\hspace{1cm}
\includegraphics[width=.45\textwidth]{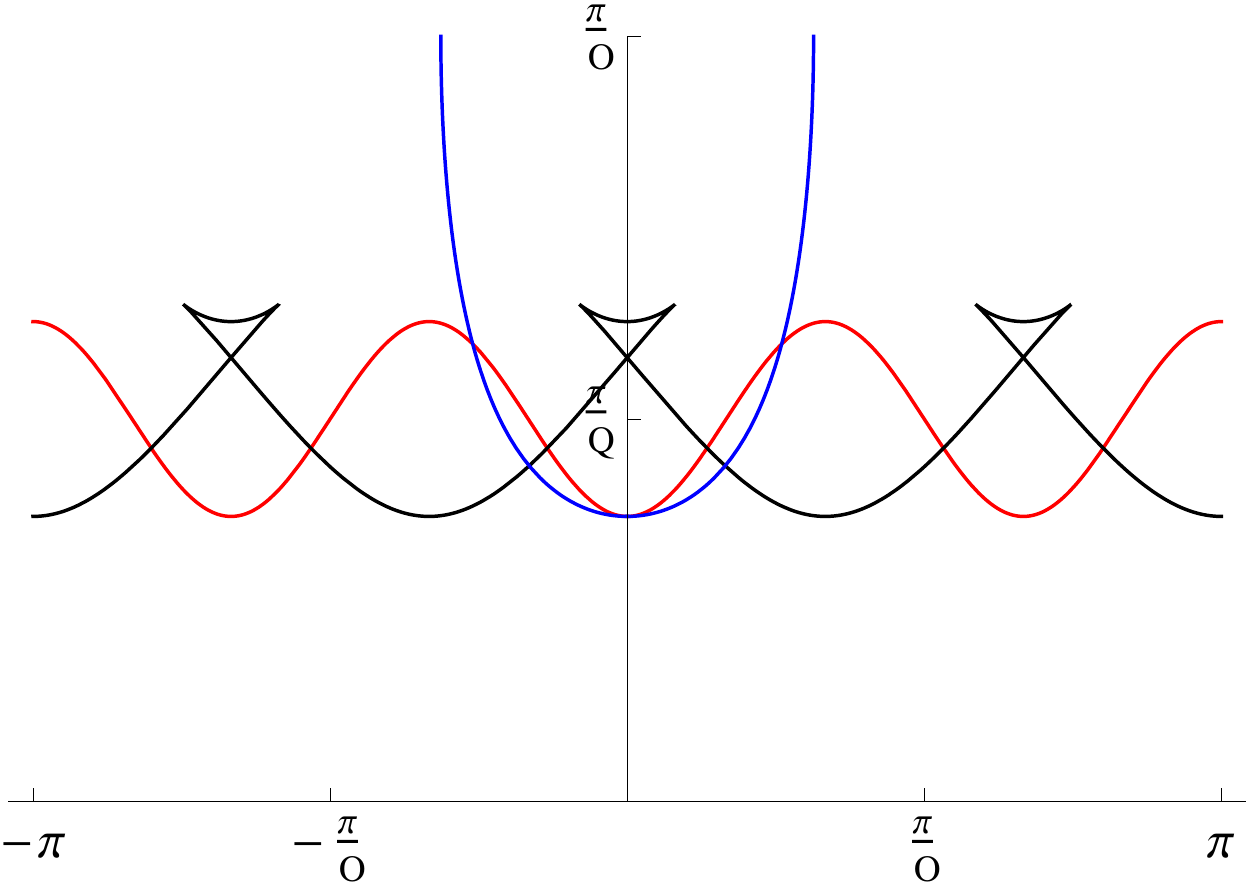}
\hspace{0.5cm}
\begin{picture}(0,0)
\setlength{\unitlength}{1cm}
\put(-0.6,0.5){$\ph$}
\end{picture}
\caption{
Similar plot as \fig{f:TS} (except that the viewpoint is shifted and the causal wedge omitted for greater ease of visualization), exemplifying how smooth $\bcrv$ can lead to kinky $\ts$.
{\bf Left:} AdS$_3$ diagram showing the bulk curve $\bcrv$ [red curve, lying at $t=0$], null normals [color-coded by starting $\ph$], which end on $\{ p_\pm(\bcpar) \}$  [black and brown cuspy curve at $r=\infty$, the black parts specifying the future/past boundaries $\tsend^\pm$ of the time strip $\ts$], and a tangent extremal surface $\extr_0$ at $\bcpar =0$ [blue curve lying at $t=0$].  For orientation, boundary $t=0$ slice is also shown [thin grey curve at $r=\infty$].   
{\bf Right:} The corresponding functions of $\ph$ describing the temporal profile of $\tbdy$ [black] which includes $\tsend$, the radial profile of $\bcrv$ [red], and the radial profile of $\extr_0$ [blue].  
Note that while $\extr_0$ is tangent to $\bcrv$ at $\bcpar=0$, it passes through $\Dhole$.
}
\label{f:TSsmC}
\end{center}
\end{figure}

Let us first consider what happens when the smooth spacelike bulk curve $\bcrv$ has a radial profile such that the corresponding $\phbdy(\bcpar)$ is not monotonic, i.e.\ $\tbdy(\ph)$ describing $\tsend$ is not single-valued.  
This is illustrated in  \fig{f:TSsmC}, which is to be directly compared to  \fig{f:TS}.  The bulk curve $\bcrv$ (red curve at $t=0$, described by smooth $\rho(\ph)$) is only slightly more wiggly than its counterpart of  \fig{f:TS}.  However, there are points (such as near $\bcpar = 0$, corresponding to red generators in  \fig{f:TSsmC}) where the expansion\footnote{
The expansion $\Theta$ of a congruence characterizes the change in the area $A(\lambda)$ along the `wavefront' with respect to the affine parameter $\lambda$, as 
$\Theta(\lambda) = \frac{1}{A(\lambda)} \, \frac{\partial}{\partial \lambda}A(\lambda)$. 
For an extremal surface, it can easily be shown \cite{Hubeny:2007xt} that the initial expansion of its normal congruence vanishes, $\Theta(\lambda =0) =0$.  Any tangent curve to an extremal surface which bends more sharply towards the boundary must then have smaller (i.e.\ negative) initial expansion.
This observation was in fact used to argue that extremal surface cannot lie inside a corresponding causal wedge \cite{Hubeny:2012wa,Wall:2012uf}.

} of these generators starts out being negative, at which points the tangent extremal surface $\extr_0$ (blue curve at $t=0$)  bends less sharply towards the boundary than does $\bcrv$.  This means that near its tangent point the extremal surface actually enters the bulk hole $\Dhole$.  Since $\extr_0$ coincides with the Rindler horizon for the $\bcpar =0$ observer, this implies that part of $\Dhole$ is causally related to part of $O_{\bcpar =0}$'s worldline.
Said differently, while it is true that the endpoints $p_\pm$ of the $\bcpar =0$ observer are lightlike-separated from the $\bcpar = 0$ point on $\bcrv$, they are timelike separated from another $\bcpar$ point on $\bcrv$.  This is directly related to the fact that the null normals emanating from $\bcrv$ intersect before reaching the boundary, which is guaranteed by the Raychaudhuri equation, as explained in \sec{s:comp}.

% Figure --------
% \fig{f:TSsmC} --

How do we then define the requisite boundary time strip $\ts$?  If we try to adopt the residual entropy rationale, we should only take the boundary points which are not in causal contact with $\Dhole$ (i.e.\ boundary points which are spacelike separated from $\bcrv$) to define the interior of the time strip $\ts$.
The time strip boundaries $\tsend^\pm$, defined by \req{tsdef}, are then subsets of the parametric curves $\{ \phbdy(\bcpar),\tbdy(\bcpar)\}$, specified by the set of  $\bcpar$-intervals such that the corresponding generators from $\bcrv$ at $\bcpar$ make it to the boundary without encountering other generators from $\bcrv$. 
This is indicated by black parts of the endpoint curves $\{ p_\pm(\bcpar) \}$ in the left panel of \fig{f:TSsmC}, while the brown parts of these curves are those endpoints $\{ p_\pm\}$ which are causally connected to $\Dhole$.  Since adjoining segments of $\tsend$ can originate from separated points $\bcpar$ on $\bcrv$, they no longer need to join on smoothly, as manifest in \fig{f:TSsmC}.

So far, we have explained why a smooth bulk curve $\bcrv$ can lead to kinky time strip boundaries $\tsend^\pm$.
But this more than a mere curiosity, since it means that our construction is no longer reversible!  There are parts of $\bcrv$ (containing, but not restricted to, those where the curvature of $\bcrv$ is such that the outgoing null expansion is negative), which do not influence $\tsend$, because the corresponding generators get cut-off by other generators.  Hence if we instead start from the corresponding time strip, i.e.\ from $\tsend^\pm$, we {\it cannot} recover the full curve $\bcrv$, but only disjoint parts of it.\footnote{
We postpone the discussion of how to obtain a complete closed bulk curve by reverse-construction till \sec{s:comp}.
}  Said differently, the construction is not one-to-one; distinct bulk curves $\bcrv$ can lead to identical boundary time strips $\ts$.   That in turn implies that if there is a boundary quantity encapsulating the residual entropy of the time strip $\ts$, and a bulk quantity encapsulating the residual entropy of the bulk hole $\Dhole$,
the two notions are distinct, and should {\it  not} be conflated.

 One of the most remarkable features of the prescription \req{diffentdiscr} of 
 \cite{Balasubramanian:2013lsa}, further explained and improved in \cite{Myers:2014jia},
  is that it remains applicable even in these types of situations, i.e.\ where a smooth bulk curve  $\bcrv$ leads to non-monotonic $\phbdy(\bcpar)$ with correspondingly multi-valued $\tbdy(\bcpar)$ and  kinky $\tsend^\pm$.  
This adds strong evidence to the basic surmise indicated above, that the quantity they construct is not really a residual entropy as intended in the motivation of  \cite{Balasubramanian:2013lsa}.

\begin{figure}
\begin{center}
\hspace{1.5cm}
\includegraphics[width=.3\textwidth]{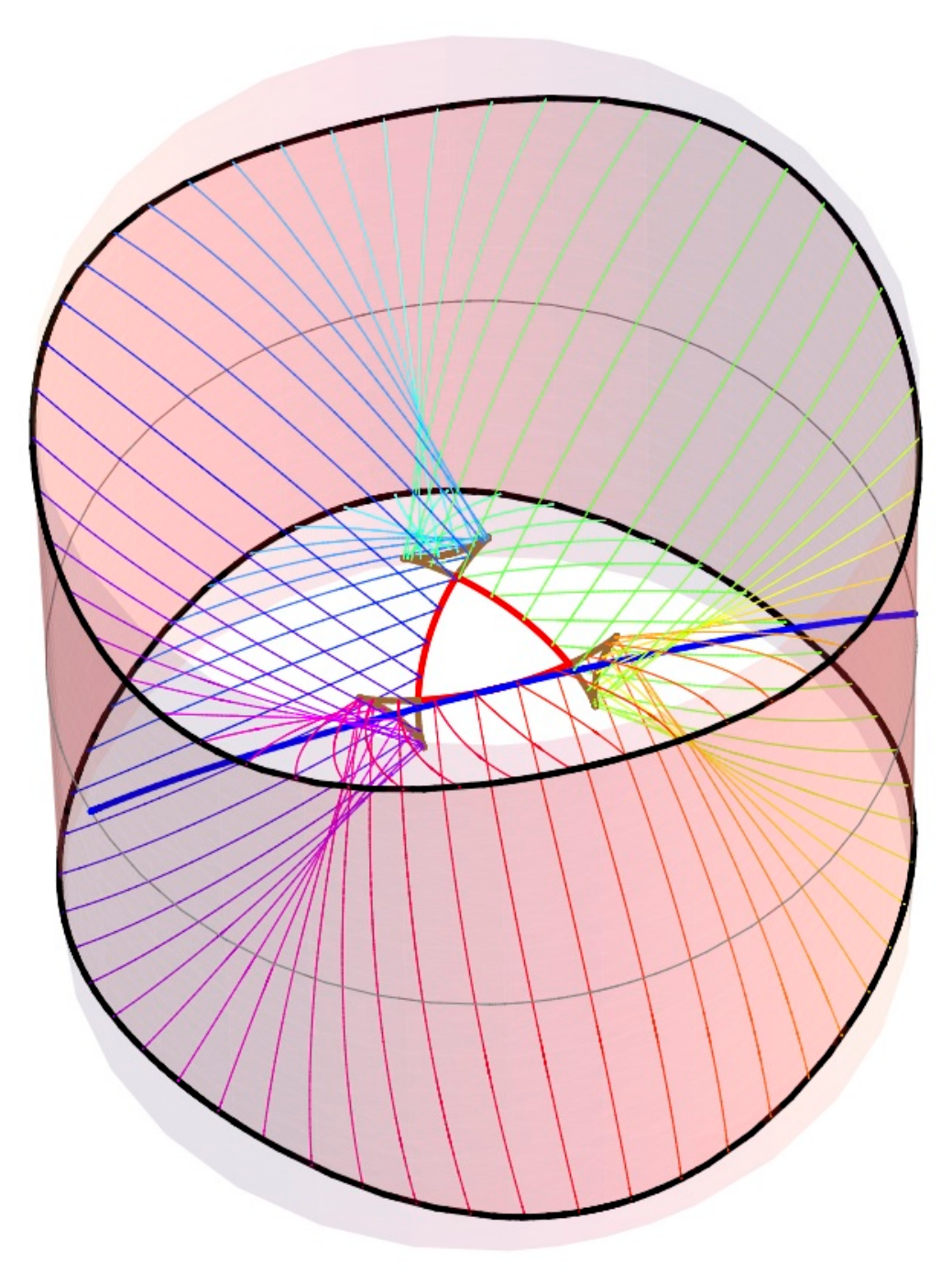}
\hspace{1cm}
\includegraphics[width=.45\textwidth]{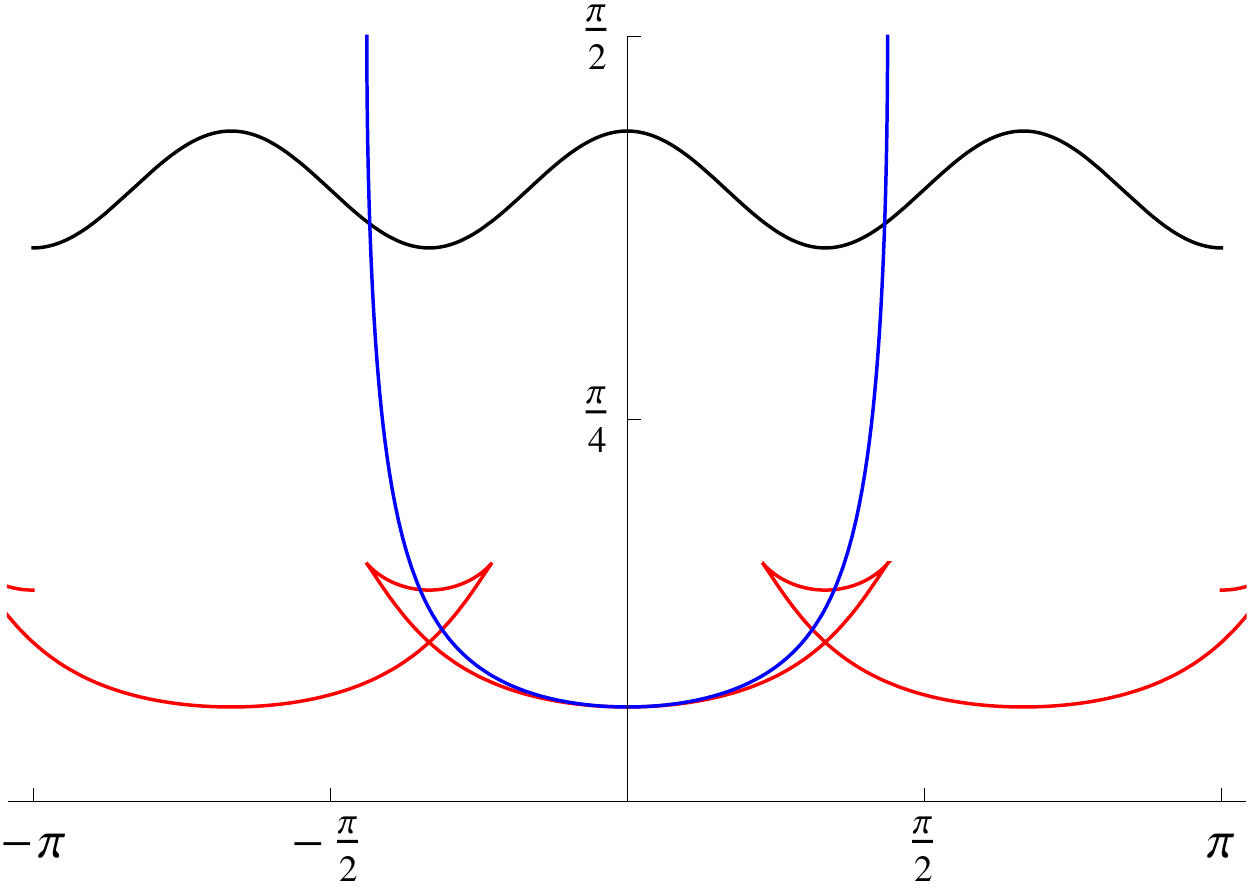}
\hspace{0.5cm}
\begin{picture}(0,0)
\setlength{\unitlength}{1cm}
\put(-0.6,0.5){$\ph$}
\end{picture}
\caption{
Similar plot as \fig{f:TS}, exemplifying how smooth $\ts$ can lead to kinky $\bcrv$ (i.e.\ a converse of \fig{f:TSsmC}).
{\bf Left:} AdS$_3$ diagram with longer boundary time strip having same wiggliness of $\tsend$ [black curves], which leads to cuspy intersection of the null normals with $t=0$ slice [red and brown curve; with only the red parts specifying $\bcrv$ and therefore $\Dhole$].  
(Similar effect could also be achieved with shorter $\ts$ but kinkier $\tsend$.)  
{\bf Right:} The corresponding temporal and radial profiles as in  \fig{f:TS} and  \fig{f:TSsmC}.
}
\label{f:TSsmSig}
\end{center}
\end{figure}

Having discussed an example wherein a smooth bulk curve $\bcrv$ leads to kinky $\tsend$, let us briefly turn to the converse manner in which the construction can fail, namely where a smooth time strip boundary $\tsend$ can lead to kinky bulk curve $\bcrv$, again rendering the construction irreversible.  A simple example is illustrated in \fig{f:TSsmSig}, which is again directly analogous to \fig{f:TS} and \fig{f:TSsmC}.  We take the time strip boundaries $\tsend$ to have precisely the same `wiggliness' (i.e.\ identical $\partial_\ph \tbdy(\ph)$) as in the smooth example of \fig{f:TS}, but the time strip $\ts$ to have overall somewhat longer duration.  (In other words, the curve $\bcrv$ in \fig{f:TS} would correspond to a constant $t\approx 0.47$ slice of the null normal congruence from $\tsend^+$ rather than the $t=0$ slice shown here.)  It is easy to see that no matter what $\tsend$ we specify, the ingoing null normals from it will eventually start to cross each other.  If they do so before reaching the $t=0$ slice, then the bulk curve $\bcrv$ they define will correspondingly have kinks, or not exist at all.  The latter happens when $\ts$ lasts long enough to render the entire $t=0$ bulk slice is causally connected to $\ts$, so that the bulk hole has closed off completely.

% Figure --------
% \fig{f:TSsmSig} --

Note that unlike the situation of \fig{f:TSsmC}, here the tangent extremal surface $\extr_0$ bends more sharply towards the boundary than $\bcrv$ (i.e.\ the null normal congruence to $\bcrv$ has positive expansion), so  $\extr_0$ remains outside $\Dhole$.  Nevertheless, the corresponding causal wedge that it rims contains some of the null generator endpoints at $t=0$ which have intersected with other generators prior to reaching $t=0$ (shown in brown in the left panel of \fig{f:TSsmSig}).  That again means that not all of the time strip boundary points $\tsend$ would be recoverable from the bulk curve $\bcrv$ (shown in red in the left panel of \fig{f:TSsmSig}) which defines the bulk hole $\Dhole$.  The only parts of $\tsend$ which can be reconstructed from $\bcrv$ are those joined by the smooth null surface generated by nowhere-intersecting null normals.  Hence the $\ts \leftrightarrow \Dhole$ mapping is neither one-to-one nor onto; distinct $\ts$'s can have the same $\bcrv$, and conversely distinct $\bcrv$'s can have the same $\ts$.
This reflects the fact that the two notions of residual entropy are distinct.

%~~~~~~~~~~~~~~~~~~~~~~~~~~~~~~~~~~~~~~~~~~~~~~~
\section{Two covariant constructions}
\label{s:constr}
%~~~~~~~~~~~~~~~~~~~~~~~~~~~~~~~~~~~~~~~~~~

Having discussed the limitations of the construction of  \cite{Balasubramanian:2013lsa}, and emphasized the main point that the differential entropy constructed therein and generalized in \cite{Myers:2014jia} is not a residual entropy as characterized in \cite{Balasubramanian:2013lsa}, we now turn to the latter.  Our basic rationale is as follows:
The motivation of \cite{Balasubramanian:2013lsa} extends well beyond the limited set-up they've analyzed.  Residual entropy might be an important and powerful construct, so it is worthwhile to try to generalize it to arbitrary regions in arbitrary asymptotically-AdS spacetimes and arbitrary dimensions, or to understand the restrictions on such a generalization.  
The motivation reviewed in the first part of \sec{s:diffentconstr} suggested a causally-defined set of relations, which can indeed be naturally defined in full generality.
In this section we set out to specify such a  construction, which is both robust and simple -- indeed the definition of the bulk regions of interest relies solely on the bulk causal structure rather than the full geometry.

As indicated in the preceding section, there are two logically distinct starting points: the boundary time strip $\ts$ which limits a boundary observers' time to make measurements, and the bulk hole $\Dhole$ which limits a bulk observer's access to a given region.  In each case, the residual entropy is supposed to encapsulate the collective ignorance of such observers.  
We will therefore present two covariant constructions, one starting from $\ts$ (representing the boundary residual entropy which we'll denote by $\REbdy$) and one starting from $\Dhole$ (representing the bulk residual entropy  which we'll denote by $\REbulk$).

%-------------------------------
\subsection{Starting from boundary time strip}
\label{s:constrbdy}
%-----------------------

Consider first the bulk construct relevant for the {\it boundary residual entropy} $\REbdy$,  specified by the collective ignorance of boundary observers due to their restriction to the time strip $\ts$.
We work in a general causal asymptotically globally AdS$_{d+1}$ spacetime $\bulk$ with boundary $\bdy$.\footnote{
\label{fn:aslocAdS}
In particular the boundary field theory lives on ESU$_d = S^{d-1} \times R$.
In fact, our construction is applicable even for more general asymptotically {\it locally} AdS spacetimes, as well as asymptotically Poincare-AdS, as discussed further in \sec{s:discuss}, 
but here we restrict attention to the simpler setup for presentational convenience.
}
We can specify a boundary time strip $\ts \subset \bdy$ as in \req{tsdef} by its future and past boundaries 
$\tsend^\pm$ 
which we take to be two non-intersecting Cauchy slices of $\bdy$
 (i.e.\ $\tsend^\pm$ are two spacelike achronal surfaces in $\bdy$ with 
$\tsend^+ \subset \Dbdy^+[\tsend^-]$,  
$\tsend^+ \cap \tsend^- = \emptyset$,
 and $\Dbdy[\tsend^\pm] = \bdy$).
Given  $\tsend^\pm$, we now define the {\it \GCW} $\CWg$ by the natural generalization of the causal wedge pertaining to the full time strip $\ts$, namely
\begin{equation}
\CWg \equiv J^+[\tsend^-] \cap J^-[\tsend^+] \ .
\label{GCWdef}
\end{equation}	
This can be thought of as the region of all bulk causal curves with both endpoints anchored on the boundary within the time strip $\ts$, and therefore it describes the bulk region which can be explored by a local bulk observer who starts and ends in $\ts$.
In fact, one could also think of $\CWg$ as the union of all individual observers' causal wedges.  Recall that any local boundary observer within the time strip has an associated causal wedge $\CWO_{p_\pm}$ defined by his worldline `endpoints' $p_\pm$ as
\begin{equation}
\CWO_{p_\pm} \equiv  J^+[p_-] \cap J^-[p_+] \ .
\label{CWOdefn}
\end{equation}	
It is then easy to show \cite{Hubeny:2013gba} that this observer's causal wedge lies within the \GCW,
\begin{equation}
\forall \ p_\pm \in \ts \ , \qquad \CWO_{p_\pm} \subset \CWg \ .
\label{}
\end{equation}	
The converse, that any point within $\CWg$ must lie in some $\CWO_{p_\pm}$, is perhaps less obvious\footnote{
E.g.\ it need not suffice to take the union of the causal wedges of just the `optimal' observers (defined as longest-lived ones for each $p_- \in \tsend^-$, and therefore having deepest-reaching causal wedges), since as discussed in the 3rd point of \sec{s:limitations}, these need not cover the entire $\ts$ (cf.\ the green-shaded region in the right panel of \fig{f:bdyTSobs}).  However, it does suffice to consider overlapping sets of optimal observers associated with each of $\tsend^+$ and $\tsend^-$ separately, and take the union of both sets of associated causal wedges.}
but is nevertheless true.

The boundary $\partial \CWg$ of this \GCW\  is comprised of two null surfaces $\cong_{\tsend^-} \subset \partial J^+[\tsend^-]$ and $\cong_{\tsend^+}\subset \partial J^-[\tsend^+]$, generated by ingoing future/past directed null geodesics emanating normally to $\tsend^\mp$, respectively.  
These generators are by definition complete towards the boundary, but they can get cut off at finite affine parameter as they penetrate into the bulk, which happens when distinct generators from the same congruence intersect.\footnote{
This set of intersections in fact separates into two types: {\it caustics}, at which neighboring generators focus and converge, and {\it crossover points}, at which non-neighboring generators cross.}
Beyond their intersection, the remainder of these null geodesics becomes timelike-separated (rather than null-separated) from its origin at $\tsend$, and therefore cannot lie on the boundary of $\CWg$.  
When this happens the null surface $\cong_{\tsend^\pm}$ is no longer smooth, but rather admits crossover seams  (generally bulk co-dimension-two sets which can degenerate, branch, etc.) terminating towards the boundary at caustic points, where neighboring geodesics intersect.
In fact, this is a rather common occurrence: not only do ordinary causal wedges exhibit such behavior as well \cite{Hubeny:2013gba}, but even black hole event horizons generically suffer from the same non-smoothness \cite{Helfer:2011vv,Beem:1997uv} (or worse -- in extreme cases they can even be `nowhere differentiable' \cite{Chrusciel:1996tw}).  Indeed, if the time strip extends over the entire boundary manifold, $\ts = \bdy$, and if the bulk contains a black hole, then the \GCW\ corresponds to the bulk domain of outer communication, and its  boundaries $\cong_{\tsend^\pm}$ precisely coincide with the event horizon.

Accepting the contingency of non-smooth boundary $\partial \CWg$ of the \GCW, let us proceed with our construction.  While the  causal set $\CWg$ defined in \req{GCWdef} is  the primary construct, given the discussion  of the previous section, it is particularly interesting to identify the analog of the bulk curve $\bcrv$  in this more general setting.  As discussed in \sec{s:limitations}, this will be a bulk co-dimension-two surface determined from $\tsend^\pm$; let us for clarity denote this surface by $\CIS$ (the subscript indicating the starting point in this construction).
Analogously to the  `causal information surface' identified in \cite{Hubeny:2012wa} as the rim of the causal wedge, it is natural to define $\CIS$ as the rim of the \GCW\ $\CWg$, given by the intersection of the null surfaces $\cong_{\tsend^\pm}$ bounding $\CWg$,
\begin{equation}
\CIS \equiv  \partial J^+[\tsend^-] \cap \partial J^-[\tsend^+] 
= \cong_{\tsend^+} \cap \cong_{\tsend^-} 
\label{gCISdefn}
\end{equation}	
Note that such surface $\CIS$ will naturally inherit the non-smoothness of the parent null surfaces, as evident in \fig{f:TSsmSig}.
It is also worth noting that  $\CIS$  does not  generically lie at a constant time (even when the spacetime itself admits a preferred notion of time due to extra symmetries).
It is easy to show that no point on $\CIS$ can be timelike-separated from $\ts$, but each point is null-separated from $\tsend^\pm$, and therefore spacelike-separated from the interior of the time-strip $\ts \backslash (\tsend^+ \cup \tsend^-)$.  
Indeed the hole $\Dhole_\tsend$ defined by $\CIS$ is precisely the set of bulk points which are spacelike-separated from interior of $\ts$; in effect $\CIS$ forms the `rim' of both the \GCW\ and the bulk hole.\footnote{
  However, as explained in the next section, these two sets -- the bulk hole and the \GCW\ -- are not constructed the same way from $\CIS$; in particular the \GCW\ generically does {\it not} coincide with the set of all spacelike-separated points to $\CIS$ external to the bulk hole.}

So far we have just used causal relations to prescribe a natural co-dimension-two bulk surface $\CIS$ determined from $\tsend^\pm$; we can now use the remaining geometrical data to obtain a characteristic number associated with this surface, namely its proper area.   In direct analogy to the `causal holographic information' $\chi$ defined in \cite{Hubeny:2012wa}, we denote this quantity by
\begin{equation}
\gchi \equiv \frac{{\rm Area}(\CIS)}{4 \, G_N}
\label{REbdyarea}
\end{equation}	
and refer to it as the `strip causal holographic information' to minimize introducing new terminology.
Note that despite the `kinks' in $\CIS$, this quantity is well-defined and finite (provided the time strip $\ts$ does not degenerate, i.e.\ $\tsend^+$ is strictly to the future of $\tsend^-$ everywhere).  

Obtaining a finite quantity, specifically the quarter-area of the corresponding bulk hole, was  presented as the key criterion of success in the differential entropy construction of  \cite{Balasubramanian:2013lsa, Myers:2014jia}.
Hence one might most naturally associate the boundary residual entropy with the strip causal holographic information $\gchi$ associated with the time strip $\ts$, i.e.\ one might be tempted to conjecture that 
\begin{equation}
\REbdy = \gchi \ .
\label{REbdyconj}
\end{equation}	
The main virtue of this proposal, apart from its simplicity and robustness, is that it indeed adheres to the expectations of   \cite{Balasubramanian:2013lsa, Myers:2014jia}; it gives a manifestly finite and well-defined quantity
which coincides with the expected bulk entanglement entropy corresponding to $\CIS$ \cite{Bianchi:2012ev}.

However, \req{REbdyconj} also has a severe drawback:  As  explained at the end of \sec{s:limitations}, different time strip boundaries $\tsend$ may lead to the same $\CIS$, and therefore the same $\gchi$; whereas it seems more natural expect that different time strip boundaries $\tsend$ should generically lead to different residual entropies $\REbdy$ -- e.g.\ we'd expect that when the boundary observers have more time for their measurements, they should generate correspondingly smaller collective ignorance.  If that is indeed the case, then the conjecture \req{REbdyconj} cannot be universally correct.  In other words, the rim surface $\CIS$ does not contain sufficient information to distinguish distinct time strips.  

In contrast, the full \GCW\ $\CWg$ {\it does} distinguish different starting strips $\ts$, i.e.\ distinct time strips $\ts$ would certainly yield distinct bulk regions $\CWg$.  Therefore we might suspect that $\REbdy$ depends more directly on some characteristic attribute of the full $\CWg$ (or perhaps its null boundary) rather than just its rim $\CIS$.
However, we may also want a quantity that does reproduce the area $\gchi$ in absence of any caustics, which poses a significant restriction on what such a quantity might be (for example, this criterion would not be fulfilled by the regularized proper spacetime volume of $\CWg$, nor the maximal spatial volume of a slice of $\CWg$, apart from the more obvious shortcoming of neither of these quantities scaling correctly with size of $\ts$).
A simpler possibility which meets this criterion, and which is in fact more reminiscent of \cite{Balasubramanian:2013lsa, Myers:2014jia}, would be to take the intersection of the null normal congruences to $\tsend^\pm$ without cutting them off at the crossover seams (e.g.\ in the left panel of \fig{f:TSsmSig} this would correspond to taking the full cuspy curve composed both of the red parts previously identified as $\bcrv$, as well as the brown parts containing the cusps).  However, this modified conjecture, too, leaves something to be desired; namely, the simple causal interpretation advocated above is no longer apparent.

In absence of further guidance, we therefore identify the full \GCW\ $\CWg$ as the requisite bulk construct from which the boundary residual entropy $\REbdy$ should be extracted, and leave the exploration of exactly {\it which} attribute of $\CWg$ to future work.
The main point we wish to emphasize here is that both $\CWg$  as well as its rim $\CIS$ are well-defined, covariant constructs, which can be uniquely determined from the given $\tsend^\pm$ in {\it any} asymptotically AdS spacetime in any number of dimensions.
Moreover, this construction did not require the extra information regarding  a requisite family of observers.  In fact, it naturally and automatically implements maximizing the region $\CWg$ over {\it all} families of observers, which in turn minimizes $\gchi$.

%-------------------------------
\subsection{Starting from bulk hole}
\label{s:constrbulk}
%-----------------------

Let us now turn to the other starting point, namely the bulk hole $\Dhole$.  This set is defined by its rim $\bcrv$, taken to be a closed spacelike achronal co-dimension-two bulk surface.   In \sec{s:diffentconstr} we have defined $\Dhole$ as the domain of dependence of a spacelike co-dimension-one region enclosed by $\bcrv$, but we can formulate an even simpler definition which dispenses with the enclosed surface (and so absolves us of having to specify a bulk Cauchy slice containing $\bcrv$).
Given $\bcrv$, we can separate the full spacetime $\bulk$ into four disjoint regions, separated by 2 null surfaces intersecting at $\bcrv$: two regions $I^\pm[\bcrv]$ composed of bulk points which are timelike-related\footnote{
By a point $p$ being timelike-related to $\bcrv$ we mean that there exists at least one point $q \in \bcrv$ which is timelike-related to $p$, i.e.\ that there exists a timelike curve between $p$ and $\bcrv$.  Conversely, the non-existence of any causal (timelike or null) curve between $p$ and any point on $\bcrv$ characterizes $p$ as spacelike-separated from $\bcrv$.  Finally, by $p$ being null-related to $\bcrv$ we mean that there is a null curve joining $p$ and some $q \in \bcrv$ while simultaneously $p$ and $\bcrv$ are not timelike-separated.
}    to $\bcrv$, and two regions which are composed of spacelike-separated points from $\bcrv$.
One of these (the compact one) is the bulk hole $\Dhole$ interior, while the other (which extends to the AdS boundary) is  (the interior of) the construct we are after.  We will dub this construct the {\it \GEW}, and denote it by $\EWg$, the subscript again indicating the starting point for the construction, in this case the bulk surface $\bcrv$ which rims $\Dhole$.

More specifically, taking the surface $\bcrv$ as our starting point, we define the \GEW\ as
\begin{equation}
\EWg = [I^+[\bcrv] \cup I^-[\bcrv]]^c \ \backslash \ \left(  \Dhole \backslash \bcrv \right)  
\label{GEWdef}
\end{equation}	
where the superscript $c$ denotes the complement of the given set (i.e.\ ${\cal S}^c \equiv \bbulk \backslash {\cal S}$).
In words, $\EWg$ is the closure of the set of  spacelike-separated points from $\bcrv$ which lie outside $\Dhole$.  (The last term in \req{GEWdef} merely ensures that $\EWg$ is a closed set which contains $\bcrv$, in analogy with $\CWg$.)  
This construction is in fact very similar to the `entanglement wedge' construction of \cite{HHLR14}:
If $\bcrv$ were an extremal surface, the rim wedge would correspond precisely to the entanglement wedge, which \cite{HHLR14} propose as the natural `dual' of the reduced density matrix specifying the entanglement entropy in question (cf.\ also \cite{Czech:2012bh}).

Once we have the \GEW, we can define the induced boundary time strip $\tsbulk$ by the restriction of $\EWg$ to the boundary,
\begin{equation}
\tsbulk = \EWg  \cap  \bdy
= [I^+[\bcrv] \cup I^-[\bcrv]]^c \ \cap \ \bdy \ .
\label{tsbulkdef}
\end{equation}	
To avoid futher new terminology, we'll refer to $\tsbulk$ simply as the `induced time strip', and we'll denote its associated boundaries by $\tsbulkend^\pm$.  Hence $\tsbulkend^\pm$ are by construction spacelike surfaces on the boundary which are null-related to the starting surface $\bcrv$.
As in the previous case of strip wedge, the \GEW\ $\EWg$ is a causal set, and as such its boundary is generated by outgoing (future and past-directed) null geodesics, emanating normally to $\bcrv$.  We denote the future and past null surfaces by 
$\cong_{\bcrv}^\pm$, so that $\bcrv = \cong_{\bcrv}^+ \cap \cong_{\bcrv}^-$ and $\tsbulkend^\pm = \cong_{\bcrv}^\pm \cap \bdy$.
As was generally the case with $\cong_{\tsend^\pm}$, the null surfaces 
$\cong_{\bcrv}^\pm$ can likewise be non-smooth; however, here the null generators are necessarily complete towards the originating surface $\bcrv$ while they need not be complete towards the boundary, which happens when they intersect other generators at finite affine parameter.  This renders $\tsbulkend^\pm$ `kinky' (as evident from \fig{f:TSsmC}).
Hence in such cases we once again see the inherent irreversibility of this construction: while a given $\bcrv$ uniquely determines the time strip $\tsbulk$,  distinct  $\bcrv$'s can potentially lead to the same $\tsbulk$.  However, it is still true that distinct  $\bcrv$'s will lead to distinct \GEW s $\EWg$.

Before proceeding to compare the two constructs further in the next section, let us briefly return to the discussion of the bulk residual entropy.
Unlike the conundrum encountered in \sec{s:constrbdy}, here we have no issue with assigning this to be the quarter-area of the surface $\bcrv$,
\begin{equation}
\REbulk = \frac{{\rm Area}(\bcrv)}{4 \, G_N}
\label{REbulkconj}
\end{equation}	
  In a sense, this is simply rephrasing the physical interpretation of Bianchi \& Myers \cite{Bianchi:2012ev}, of the bulk entanglement entropy of the hole as the residual entropy pertaining to observers who lack causal access to the hole.
The interesting question, which we again leave for the future, is how to recover this from the boundary data.  If we only have access to the time strip $\tsbulk$, then as pointed out above, we will not be guaranteed to recover $\bcrv$ and therefore  $\REbulk$ given by \req{REbulkconj} from this data.  In fact, this appears to be a more serious worry than the converse one of  \sec{s:constrbdy}, since in the latter we assumed that a bulk observer has access not only to $\CIS$ but any geometrical construct in the bulk.  In contrast, a boundary observer does not have any apparent access to the bulk $\EWg$ but only to its boundary restriction $\tsbulk$.

Given this, we can make similarly open-ended remarks as in \sec{s:constrbdy}:  we do not propose a definite boundary prescription of how to obtain a number specifying the bulk residual entropy from boundary data.  Instead, we present a new set of constructs, $\EWg$ and corresponding $\tsbulk$, which are equally robust as the concept of residual entropy and capture the essence of its meaning.  They are well-defined sets in any asymptotically AdS spacetime, in any dimension, and for any bulk hole.

%~~~~~~~~~~~~~~~~~~~~~~~~~~~~~~~~~~~~~~~~~~~~~~~
\section{Wedge comparison}
\label{s:comp}
%~~~~~~~~~~~~~~~~~~~~~~~~~~~~~~~~~~~~~~~~~~

Let us now take stock of the two constructs we've presented in the previous section, namely the \GCW\ $\CWg$ (defined by \req{GCWdef}, given a time strip bounded by $\tsend^\pm$) and the \GEW\ $\EWg$  (defined by  \req{GEWdef}, given a bulk hole, rimmed by $\bcrv$). 
Both are fully covariantly defined bulk co-dimension-zero regions which don't require any special symmetries for their construction.
Motivated by the essence of the concept of residual entropy as collective ignorance, both regions are causal sets, and hence are in a sense the most elemental geometrical constructs.
Since causal sets have boundaries generated by null geodesics, one may wonder what exactly is the distinction between the \GCW\ and the \GEW.   This section examines the relation between these two constructs in greater depth.

% Figure 
\begin{figure}
\begin{center}
\includegraphics[width=.45\textwidth]{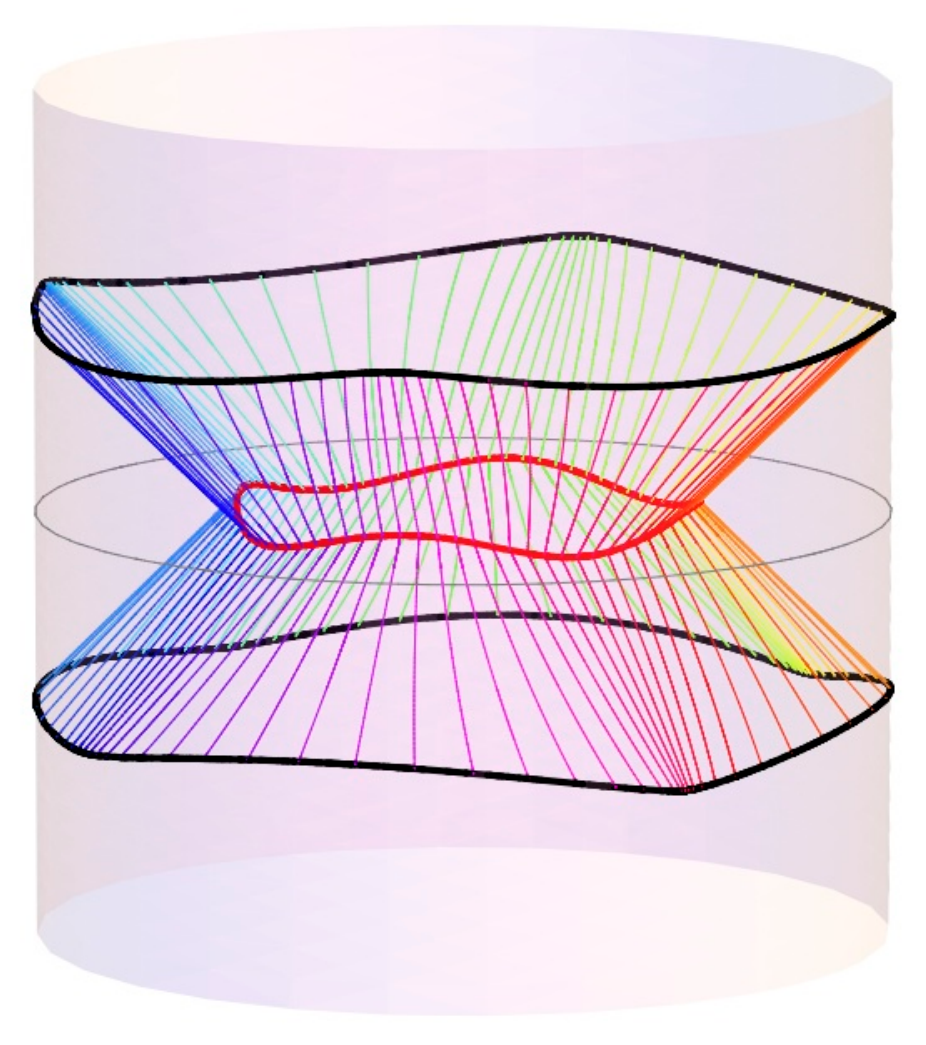}
\caption{Example of a `tame' setup, wherein a smooth bulk rim $\bcrv$ [thick red curve] yields smooth time strip $\tsend^\pm$  [thick black curves] and vice-versa.  In this case the \GCW\ $\CWg$ and the \GEW\ $\EWg$ coincide, and the null generators [thin curves color-coded by $\phc$]  are normal to both $\bcrv$ and $\tsend^\pm$.
Note that both $\bcrv$ and $\tsend^\pm$ vary in $t$; no symmetries are present in either $\bcrv$ or $\tsend^\pm$.
 However, this `genericity' is rather fragile: even innocuously small deformations to either starting point can easily destroy the smoothness.  The greater the separation between $\bcrv$ and $\tsend^\pm$, the less wiggliness is admissible in order for this tameness to be maintained.}
\label{f:CWEWsmooth}
\end{center}
\end{figure}
% \fig{f:CWEWsmooth}

As already mentioned above, if the setup is sufficiently `tame' so as to allow the null generators in each congruence to connect the bulk hole rim $\bcrv$ to boundary time strip ends $\tsend^\pm$ without intersecting other generators, then the null surface forming the wedge boundary is smooth, the two wedges coincide, and each construction is reversible.  A time-flip-symmetric example of such a situation was already shown in  \fig{f:TS}; but a more general example is illustrated in  \fig{f:CWEWsmooth}, wherein both the bulk surface $\bcrv$ and the time strip ends $\tsend^\pm$ are smooth spacelike surfaces and we can clearly see that the generators in a given congruence do not intersect each other.
 
Let us now examine the reversibility of our construction.  If we start from $\bcrv$ and construct an outgoing future-directed family of null normal geodesics (let us call this congruence $\cong_\bcrv^+$), the endpoints define a boundary slice $\tsend^+$, and similarly the endpoints of outgoing past-directed null congruence (denoted $\cong_\bcrv^-$) define $\tsend^-$.  
Having thus determined  the  boundary slices $\tsend^\pm$, we can take these as our starting point, and  consider ingoing past/future-directed null normals (labeled by $\cong_{\tsend^\pm}$, the subscript specifying  which surface the congruence emanates from), following each till they intersect.  The statement of reversibility is that  
$\cong_{\tsend^+} = \cong_\bcrv^+$ and $\cong_{\tsend^-} = \cong_\bcrv^-$, so in particular the original bulk surface $\bcrv$ lies at the intersection of these two congruences, $\bcrv = \cong_{\tsend^-} \cap \cong_{\tsend^+}$.  

The underlying reason for this reversibility of construction is that given a smooth null hypersurface $\cong$ generated by a congruence of null geodesics (let us denote their tangent vector by $k^a$), any vector $\xi^a$ tangent to $\cong$ must be normal to $k^a$, i.e.\ $k_a \, \xi^a = 0$.  Hence the null generators will stay normal  to {\it any} spatial slice of the null hypersurface $\cong$; here $\bcrv$ and $\tsend$ are simply specific realizations of such spatial slices.  However, this reasoning breaks down if ${\cal N}$ ceases to be smooth, which happens whenever any of its generators intersect.  

We have seen two examples of such an occurrence in \sec{s:limitations}; cf.\ \fig{f:TSsmC} and  \fig{f:TSsmSig}.
When two generators of the same congruence intersect, they cease to generate the boundary of the causal set beyond their intersection.  The null hypersurface $\cong$ forming the boundary of such causal set then admits caustics and crossover seams, which induces kinks in any slice of $\cong$ beyond the earliest caustic.

In case of $\cong_\bcrv^\pm$, the presence of caustics is related to the behavior of the null expansion $\Theta$  along congruence from $\bcrv$, as follows.
Consider the Raychaudhuri equation along any generator $k^a$, measured with respect to some affine parameter $\lambda$, in $d+1$ spacetime dimensions
\begin{equation}
\frac{d \Theta}{d\lambda} = 
- \frac{1}{d-1} \, \Theta^2 - \sigma_{ab} \, \sigma^{ab} 
- R_{ab} \, k^a \, k^b
\label{Raych}
\end{equation}	
where the first two terms on the RHS relate to the square of expansion and shear (the vorticity vanishing by hypersurface orthogonality),  and the expansion 
$\Theta = \nabla_{\! a} \, k^a$ can be thought of as differential change in the area element
\begin{equation}
\Theta(\lambda) = \frac{1}{A(\lambda)} \, \frac{\partial}{\partial \lambda}A(\lambda) \ .
\label{Thetadef}
\end{equation}	
The null energy condition guarantees that $R_{ab} \, k^a \, k^b \ge 0$, which implies $\frac{d \Theta}{d\lambda} \le 0$, so that once $\Theta(\lambda)$ becomes negative it must diverge, $\Theta(\lambda \to \lambda_\wedge) \to -\infty$ at some finite affine parameter $\lambda_\wedge$, signalling the presence of a caustic. 
Since the boundary lies at infinite $\lambda$, such a generator does not reach the boundary, being cut off by neighboring generators.  Nearby generators along this congruence then likewise intersect each other.
 In fact, the crossover seam which marks the intersection of non-neighboring generators occurs before the latter encounter caustics, i.e.\ for some $\lambda < \lambda_\wedge < \infty$ (in case of the future-directed congruence).  Once formed, a crossover seam continues all the way to the boundary, causing a kink in the corresponding $\tsbulkend$. 

In pure AdS$_3$, the criterion for both presence and absence of caustics becomes particularly simple, since the Raychaudhuri equation simplifies dramatically: the congruence (parameterized by $\bcpar$) is one-dimensional, so the shear and expansion automatically vanish, and the Ricci tensor contracted with the tangent vector to null geodesics likewise vanishes since $R_{ab} = \Lambda \, g_{ab}$, so we have
\begin{equation}
\frac{d \Theta}{d\lambda} = - \Theta^2 
\qquad \Rightarrow \qquad
\Theta(\lambda) = \frac{\Theta_\bcrv}{1+ \Theta_\bcrv \, \lambda}
\label{}
\end{equation}	
where $\Theta_\bcrv \equiv \Theta(\lambda_\bcrv)$ is the initial expansion at $\bcrv$.
This means that if the initial null expansion $\Theta_\bcrv(\bcpar) < 0$ for {\it any} $\bcpar$ along $\bcrv$, then the corresponding generators develop caustics, rendering $\tsbulkend$ kinky.  In this case, some generators corresponding to  $\bcpar$ for which $\Theta_\bcrv>0$ may also intersect other (non-neighboring) generators.  However, if $\Theta_\bcrv(\bcpar)>0$ for {\it every} $\bcpar$ along $\bcrv$, then (in pure AdS$_3$) we cannot achieve intersections amongst the outgoing generators, so the induced time strip ends  $\tsbulkend^\pm$ remain smooth.

In more general spacetimes, however, the smoothness criterion is more complicated since it is determined by the full spacetime geometry along $\cong_\bcrv$ via \req{Raych} rather than just locally at $\bcrv$.  In particular, $\Theta_\bcrv>0$ everywhere along $\bcrv$ does not guarantee smoothness of $\tsbulkend$.  For instance, in Schwarzschild-AdS spacetime, a spherical closed surface $\bcrv$ lying outside the black hole (not wrapping it) has everywhere positive expansion $\Theta_\bcrv$, whilst the corresponding $\tsbulkend^\pm$ necessarily has a kink at the axis of symmetry due to intersections of generators which pass from around opposite sides of the black hole.

For the \GCW\ null congruences 
$\cong_{\tsend^\pm}$, the situation is even more complicated, since smoothness of $\CIS$ depends not only on the shape of $\tsend^+$ and $\tsend^-$, but also on the time separation between them, i.e.\ the width of the time strip $\ts$;  wider  $\ts$ means that  less time-variation in $\tsend^\pm$ is admissible in order to maintain smooth $\CIS$.   For example, for pure AdS, as $\tbdy^\pm \to \pm \frac{\pi}{2}$, the bulk hole closes off at a point where all generators intersect, so arbitrarily small variation is $\tbdy(\ph)$ around this value would lead to a caustic prior to reaching the center.  Conversely, in any smooth asymptotically-AdS spacetime, given arbitrarily wiggly smooth spacelike $\tsend^-$, we can always find a smooth $\tsend^+$ which is sufficiently near to render the corresponding $\CIS$ smooth.

\begin{figure}
\begin{center}
\hspace{1.cm}
\includegraphics[width=.35\textwidth]{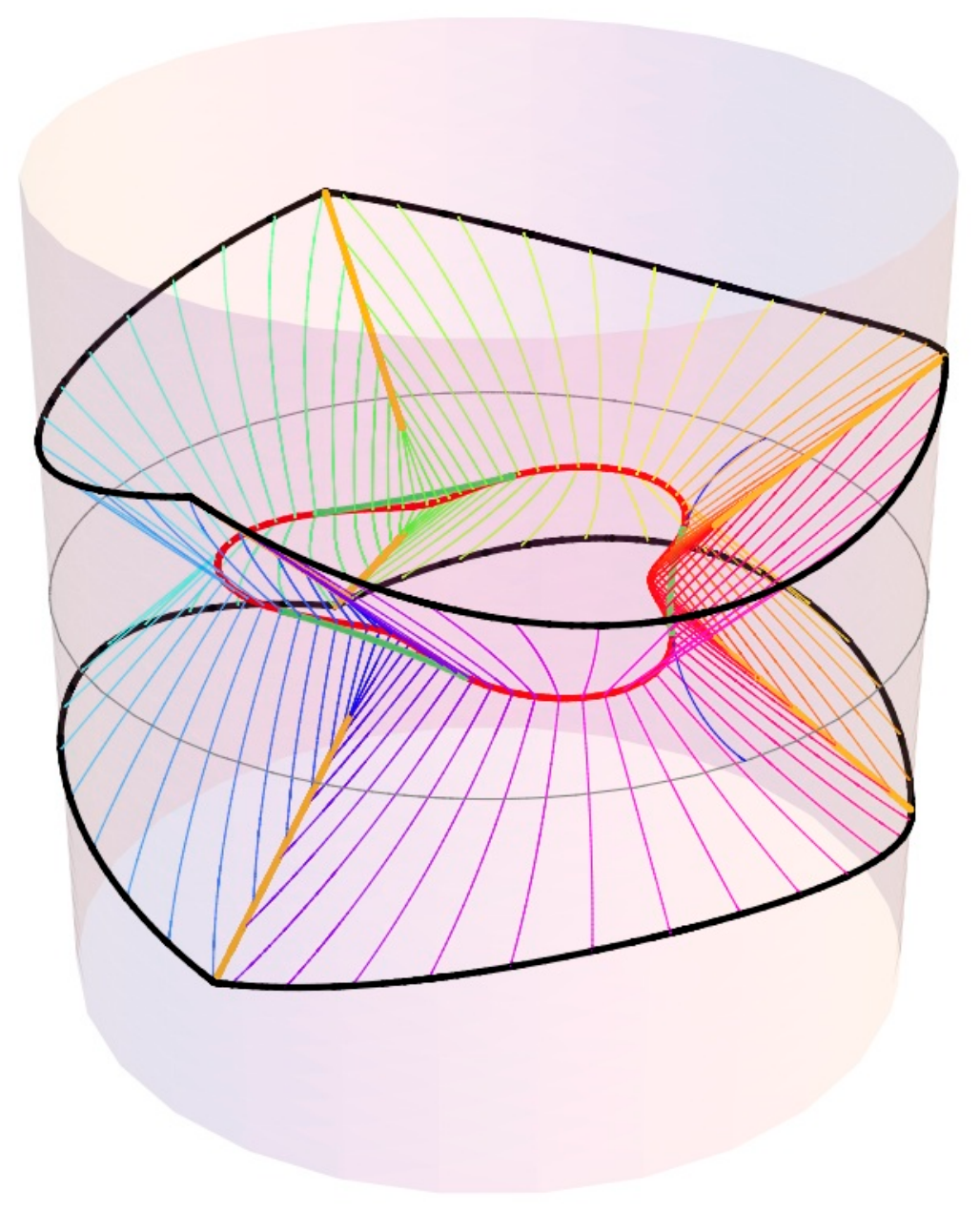}
\hspace{1.5cm}
\includegraphics[width=.35\textwidth]{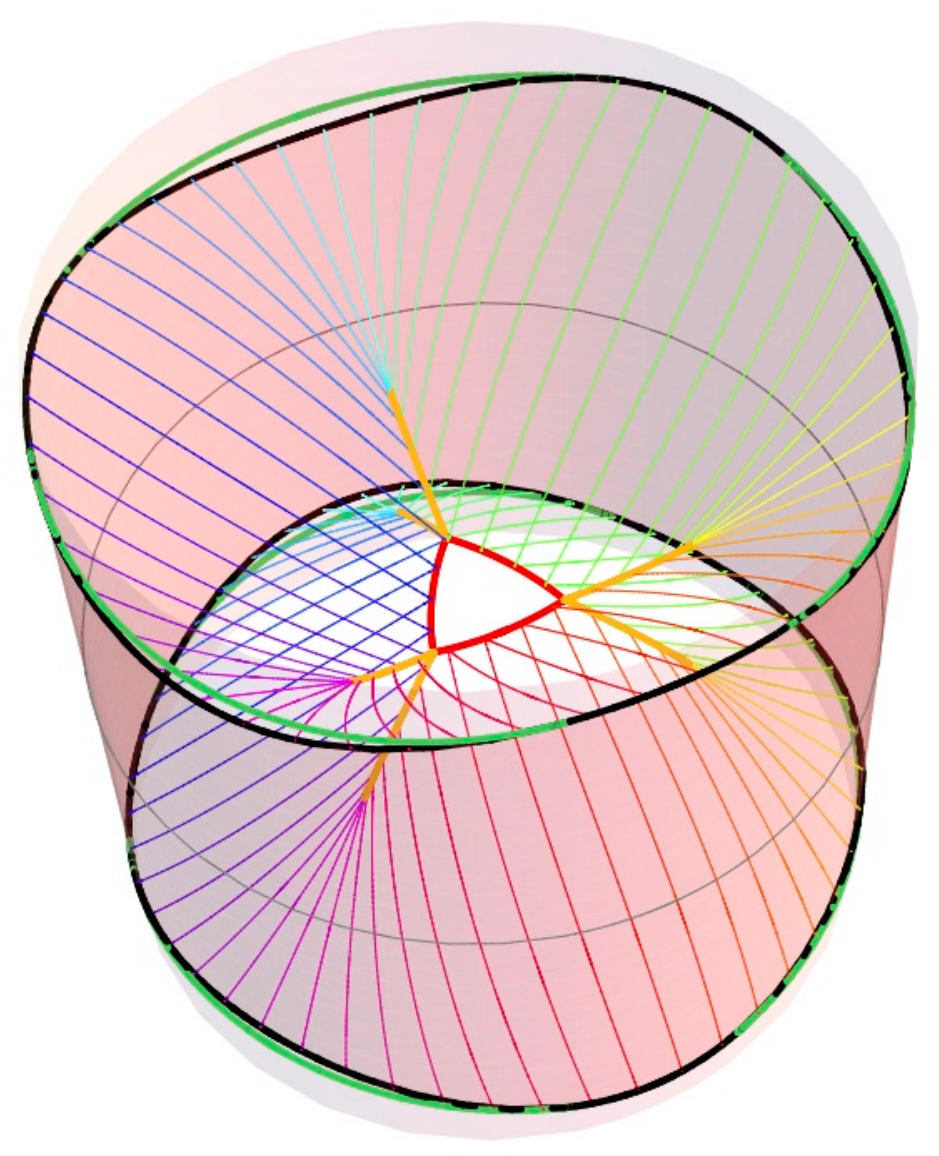}
\caption{The two covariant constructions discussed in \sec{s:constr},
directly constructed from the examples of \fig{f:TSsmC} and  \fig{f:TSsmSig} (therefore the two panels are unrelated to each other; we discuss the reverse-construction for each example separately.)  
{\bf Left:} The \GEW\ $\EWg$ 
discussed in \sec{s:constrbulk}, bounded by null surfaces $\cong_{\bcrv}^\pm$ constructed from null normals to $\bcrv$ [thick red curve].  The corresponding crossover seams [thick orange curves] continue to the boundary, where they give rise to kinks in $\tsbulkend^\pm$ [thick black curves].  If we instead start from the kinky  $\tsbulkend^\pm$ and construct the corresponding \GCW, we recover only part of $\bcrv$; the differing parts are given by segments of causal information surface for each kink [thick green curves], which is tangent to $\bcrv$ [the full such surface for the $\bcpar = 0$ kink is indicated by thin blue curve].
{\bf Right:} The \GCW\ $\CWg$ 
discussed in \sec{s:constrbdy}, bounded by null surfaces $\cong_{\tsend^\pm}$ generated by null normals to $\tsend^\pm$ [black].  The corresponding crossover seams [orange] continue to the bulk rim $\CIS$ [red], generating kinks in the latter.  If we reverse the construction and generate \GEW\ from kinky $\CIS$, we recover only part of the original $\tsend^\pm$, the differing parts again indicated by the green segments on the boundary.
}
\label{f:EWCWirreversible}
\end{center}
\end{figure}

Having discussed the criteria for tameness of the setup (by which we mean smoothness of both $\bcrv$ and $\tsend^\pm$, and therefore reversibility of the construction, $\CWg = \EWg$), let us now examine the general non-tame setup a bit further.  
Given that $\CWg \ne \EWg$ whenever either congruence admits caustics, we wish to explore what happens when we try to reverse the construction.  
For the two examples presented in \sec{s:constr} (namely  \fig{f:TSsmC} and  \fig{f:TSsmSig}), we illustrate the \GEW\ and \GCW, along with the corresponding reverse construction, in left and right panels of \fig{f:EWCWirreversible}, respectively.  

% Figure 
% \fig{f:EWCWirreversible}

Note that the kinks have a definite orientation.  The kinks in $\tsbulkend^\pm$ `point' away from the induced time strip $\tsbulk$ (so that the tangent to $\tsbulkend$ from either side of the kink leaves $\tsbulk$), whereas the kinks in $\CIS$ point away from the bulk hole $\Dhole$.  When reversing the construction, in both cases the corresponding causal set therefore contains a finite part of the light cone from the kink point, rather than just a single null geodesic.  To see this, one can either use the definition directly, to one can take a limiting procedure of considering a family of smooth surfaces with kinked limit.  

The above observation immediately lets us formulate a relation between the \GEW\ and the \GCW.
A spacelike surface corresponding to the crossover seam which terminates at the kink lies outside the light cone from the kink.
Therefore the null generators terminating on the crossover seam will likewise lie on one side of the kink  light cone.  In case of the \GEW\  (cf.\ left panel of \fig{f:EWCWirreversible}) the crossover seam lies outside a light cone from the kink in $\tsbulkend$, so the reverse-constructed \GCW\ is contained within the original \GEW, $\CWg \subset \EWg$, and correspondingly the induced rim $\CIS$ (containing the green segments in \fig{f:EWCWirreversible})  lies closer to the boundary than the original $\bcrv$.  Conversely, in the case of the \GCW\  (cf.\ right panel of \fig{f:EWCWirreversible}) the crossover seam lies outside the light cone from the kink of $\CIS$, which means that the resulting time strip induced by the reverse-constructed \GEW\ from $\CIS$ contains the original time strip (again, the difference is indicated by the green segments in \fig{f:EWCWirreversible}).  Correspondingly, the reverse-constructed  \GEW\ contains the \GCW, so $\EWg \supset \CWg$.

Said more formally, any  point $p \in \CWg$ is spacelike related to the rim $\CIS$, and therefore by definition $p \in \EWg$.  However, since in presence of crossover seams there exits points $p \in \EWg$ which are spacelike-related to $\tsbulkend$, such $p \notin \CWg$.
This argument shows that $\CWg \subset \EWg$ in full generality, namely the \GCW\ pertaining to boundary observers is contained in the corresponding \GEW\ pertaining to bulk observers.  

%~~~~~~~~~~~~~~~~~~~~~~~~~~~~~~~~~~~~~~~~~~~~~~~
\section{Discussion}
\label{s:discuss}
%~~~~~~~~~~~~~~~~~~~~~~~~~~~~~~~~~~~~~~~~~~

The concept of entanglement entropy associated with some subsystem in a quantum theory is verily an important one.  Over the last decade, it has been extensively and fruitfully explored in the context of holography, bolstering the intriguing expectation that natural quantum information theoretic quantities may carry deep connections to geometry.
Although entanglement entropy is typically formulated in terms of reduced density matrix for the given subsystem (obtained by tracing over the degrees of freedom associated with its complement), it can be used to quantify the ignorance of an observer who only has access to the said subsystem.

In contrast, the notion of {\it collective} ignorance, pertaining to some specified {\it family} of observers, is much less familiar. The authors of \cite{Balasubramanian:2013lsa} attempted to develop this notion in holography in the case of observers who lack access to a certain spacetime region, initially coining the term `residual entropy' to describe this collective ignorance. 
More specifically,  
the unknown region for bulk observers
is a spacetime `hole' $\Dhole$, which \cite{Balasubramanian:2013lsa} then recast in terms of boundary observers restricted to a finite `time strip' $\ts$; this motivated a formula for the collective ignorance in terms of a differential expression involving entanglement entropies, which was recently extended by \cite{Myers:2014jia} who renamed this construct `differential entropy'.

However, as explained in detail in \sec{s:limitations}, this differential entropy construction of \cite{Balasubramanian:2013lsa,Myers:2014jia} does {\it not} generally adhere to the notion of residual entropy (even in the rather limited context of pure AdS$_3$ bulk spacetime, and even for time-flip-symmetric holes or strips); rather, it appears to be a genuinely distinct notion.  Nevertheless, the residual entropy (defined as indicated above, in terms of a collective ignorance) is an intriguing concept, which may well be useful as a more refined measure of where/how the information about the system is stored.  This motivated the main point of the present work, which proposes a robust, well-defined bulk construction based on the essence of residual entropy, which does not suffer from any of the limitations discussed in \sec{s:limitations}.  In fact,  as mentioned in footnote \ref{fn:aslocAdS}, our residual entropy constructions are even generalizable to asymptotically locally AdS$_{d+1}$ (rather than just asymptotically globally AdS$_{d+1}$) spacetimes, with arbitrary\footnote{
Of course, we are restricting attention to causal Lorentzian boundary geometries.  However, the metric need not solve any field equations (there is no dynamical gravity), nor does it need to be causally trivial, e.g.\ the boundary can admit black holes (as for the bulk black funnel/droplet spacetimes introduced in \cite{Hubeny:2009ru}).
} boundary metric, as well as to more general theories of gravity which admit causal constructs.

In \sec{s:constr} we have identified two covariantly defined bulk co-dimension-zero regions:  The \GCW\ $\CWg$ (cf.\ \req{GCWdef}) is the set of causally-connected (both in future and past direction) points to the boundary time strip $\ts$, while the \GEW\ $\EWg$  (cf.\ \req{GEWdef}) is the closure of the set of spacelike-separated points from the bulk hole $\Dhole$.  These provide natural geometrical quantities associated with boundary and bulk residual entropies, respectively.  While \cite{Balasubramanian:2013lsa} implicitly conflated the bulk and boundary notions of collective ignorance, we have seen that the corresponding wedges coincide only for sufficiently tame situations (such as illustrated in \fig{f:CWEWsmooth}); more generally they are distinct, despite both being generated by bulk null geodesics. 

In \sec{s:comp} we explained this distinction and offered a simple criterion for its existence.  In particular, the presence of caustics  in the null generators of the wedge boundary renders the latter non-smooth at the corresponding crossover seams, which results in kinky $\tsbulkend$ from smooth $\bcrv$, or conversely, kinky $\CIS$ from smooth $\tsend$.
Both of these situations are exemplified in \fig{f:EWCWirreversible}, the left and right panels presenting \GEW\ and \GCW, respectively.  In each case, reversing the construction yields a surface which is distinct from the starting point (indicated by the green segments in \fig{f:EWCWirreversible}).
Nevertheless, in both cases, we see the inclusive relation $\CWg \subset \EWg$, namely the \GCW\ pertaining to boundary observers is contained in the corresponding \GEW\ pertaining to bulk observers.
 This relation is a generalization of the statement derived in \cite{HHLR14} (cf.\ also the earlier works \cite{Hubeny:2012wa,Wall:2012uf}) that the causal wedge is contained in the entanglement wedge.
 
One lesson from our construction is that working directly with a family of local observers, while conceptually a useful crutch, is rather more awkward than working directly with causal relations.  We have seen that although the envelope of individual causal wedges does reproduce the full causal wedge, combining the individual causal holographic information quantities pertaining to each observer does not seem to work when used analogously to the differential entropy construction of \cite{Balasubramanian:2013lsa}.  Rather, it is the (in)accessibility of a given region which specifies the residual entropy.
Moreover, observer characterization is not unique; even for a set of `optimal' observers discussed in \sec{s:limitations}, only the endpoints of observers' worldlines matter, so specifying  the entire observer trajectories constitutes unnecessary information, unless one relies on a more detailed definition specifying the given observers' measurements etc.
In this regard, a definition along the lines of Kelly \& Wall's proposal relating causal holographic information to one-point entropy   \cite{Kelly:2013aja}
seems more natural; however, it is somewhat unclear how specifying all one-point functions in $\ts$ relates to specifying the observations of a family of observers restricted to $\ts$.  
More generally, it is fair to ask whether the notion of collective ignorance really is a well-defined one, and for this discussion it would certainly be useful to have a sharp operational definition of what one means by it.  

Let us now revisit the intriguing issue of how do our causal constructs $\CWg$ and $\EWg$ actually relate to residual entropies $\REbdy$ and $\REbulk$.  As argued in \sec{s:constr}, the most natural guess given by the rim quarter-area, \req{REbdyconj} and \req{REbulkconj}, is on further reflection rather perplexing.  In particular, in the general case involving caustics of null generators of the wedge boundary, the consequent irreversibility of the construction implies that different boundary time strips $\ts$ can lead to the same $\gchi$, and conversely different-area bulk holes can lead to the same time strip $\tsbulk$.

In the former case, wherein we wish to describe the boundary residual entropy $\REbdy$ in terms of bulk constructs, we have two  natural resolutions to this tension:
First, as suggested above, the residual entropy might actually {\it not} be given by the area of the induced rim \req{REbdyconj}, but rather by some other quantity pertaining to the full \GCW\ $\CWg$, since the latter does have a one-to-one relation with the defining boundary time strip $\ts$.  In this case, though, such quantity remains to be identified and justified.  Second, if \req{REbdyconj} prevails, it has the curious implication that the collective ignorance is a more globally defined notion than one given by merely composing individual observers' ignorance: in particular, it would imply that for some of the observers, having an extra bit of time to make measurements makes no difference for the collective ignorance.  For example, in the case presented in the right panel of \fig{f:EWCWirreversible}, the boundary observers having time up to the green curve (describing the reconstructed $\tsbulk$ from $\CIS$) would amount to same collective ignorance as the shorter-lived set of boundary observers restricted to the original $\ts$ indicated by the black curves.

Indeed, the necessity of accounting for global relation amongst the observers is apparent also in the context of the observation made earlier, that if we take the boundary time strip $\ts$ to have long enough duration, there is no causally disconnected bulk hole; in which case  \req{REbdyconj} would imply that the boundary residual entropy vanishes.  Nevertheless, one might have a-priori expected that a finite, rather than infinite, duration time strip means that there is still some uncertainty about the full state of the system, reflecting the fact that observers can make measurements for only a finite time.\footnote{
  Indeed, \cite{Balasubramanian:2013lsa} mentioned this puzzle in relation to the energy-time uncertainty relation.  }
  If we had a definitive argument that the boundary residual entropy must remain nonzero for any finite-duration time strip, then \req{REbdyconj} would be manifestly ruled out.  However in absence of such an argument, either of the alternatives mentioned above remains a possibility.
It is also interesting to note the curious asymmetry between the bulk and boundary residual entropy concepts: In the bulk, we're leaving out only a finite spacetime volume and keeping the entire (infinite) rest of the space, whereas in the boundary we're leaving out infinite time intervals and keeping only a finite-duration strip.  Nevertheless, we expect that both bulk and boundary residual entropies should be finite quantities.

Let us now return to the converse puzzle, of how to describe the bulk residual entropy in terms of boundary quantities.  This case  in fact appears to be the  more problematic one, since we have not identified any alternate natural boundary quantity related to the bulk hole, besides the time strip.  In absence of such a construct, one is led to a speculation that {\it  no} set of local boundary observers can recover the bulk residual entropy given by 
\req{REbulkconj} (which coincides with the bulk entanglement entropy of $\bcrv$ proposed by \cite{Bianchi:2012ev}).  
Nevertheless, the potential necessity of non-local specification of the set of boundary observers whose `collective ignorance' we wish to capture does not necessarily invalidate the original motivation of \cite{Balasubramanian:2013lsa}.  The standing challenge, however, is to formulate a more precise definition of the concept of residual entropy.
These speculations illustrate the subtleties in trying to formulate holographic measures of quantum information.

%~~~~~~~~~~~~~~~~~~~~~~~~~~~~~~~~~~~~~~~~~~~~~~
\acknowledgments 
%~~~~~~~~~~~~~~~~~~~~~~~~~~~~~~~~~~~~~~~~~~~~~~
It is a pleasure to thank Jan de Boer, Roberto Emparan, Alberto Guijosa, Matt Headrick, Michal Heller, Juan Maldacena, Rob Myers, Mukund Rangamani, Simon Ross, Masaki Shigemori, and Tadashi Takayanagi for useful discussions.  I would also like to thank the Isaac Newton Institute, Aspen Center for Physics,  Institute for Advanced Study, and  Yukawa Institute for Theoretical Physics for hospitality during this project.
This work was supported by the STFC Consolidated Grant ST/J000426/1, FQXi Grant RFP3-1334, and the Ambrose Monell Foundation.

\appendix

%~~~~~~~~~~~~~~~~~~~~~~~~~~~~~~~~~~~~~~~~~~~~~~~
\section{Glossary of notation}
\label{s:notation}
%~~~~~~~~~~~~~~~~~~~~~~~~~~~~~~~~~~~~~~~~~~

Since there are many related  constructs we discussed, and in some cases we invented the notation and terminology for them, here we present an overview for ease of orientation.
This is summarized in Table \ref{t:glossary}, where we list the notation, terminology/description, dimensionality for asymptotically AdS$_{d+1}$ bulk, and defining relation for the most important constructs.

\begin{table}[htdp]
\caption{Glossary of notation and terminology for causal constructs.  The heading abbreviations in the last two columns refer to dimensionality of the given construct and link to the defining relation.}
\begin{center}
\begin{tabular}{|c|c|c|c|}
\hline \hline
notation & terminology/description & dim. & defn. 
\\ \hline \hline
% time strip
$\ts$ & boundary time strip & $d$ & \req{tsdef} 
\\ \hline
% time strip boundaries
$\tsend^\pm$ & future/past time strip boundaries & $d-1$ & above  \req{GCWdef} 
\\ \hline
% null congruences from bdy
$\cong_{\tsend^\pm}$ & null geodesic congruences from $\tsend^\pm$ & $d$ & %below \req{GCWdef}
 above \req{gCISdefn}
\\ \hline
% generalized causal wedge
$\CWg$ & \GCW & $d+1$ & \req{GCWdef} 
\\ \hline
% generalized CIS
$\CIS$ & rim of the \GCW & $d-1$ & \req{gCISdefn} 
\\ \hline
% generalized chi
$\gchi$ & strip causal holographic information & $0$ & \req{REbdyarea} 
\\ \hline
% bdy Residual entropy
$\REbdy$ & boundary residual entropy & $0$ &
\req{REbdyconj} and below
\\ \hline \hline
% bulk hole
$\Dhole$ & bulk hole & $d+1$ & early \sec{s:constrbulk} 
\\ \hline
% bulk curve/surface/rim of hole
$\bcrv$ & bulk hole rim & $d-1$ & early \sec{s:constrbulk}
\\ \hline
%  null congruences from bulk
$\cong_{\bcrv}^\pm$ & null geodesic congruences from $\bcrv$ & $d$ & below \req{tsbulkdef} 
\\ \hline
% generalized entanglement wedge
$\EWg$ & \GEW & $d+1$ & \req{GEWdef} 
\\ \hline
% induced time strip
$\tsbulk$ & induced time strip & $d$ & \req{tsbulkdef} 
\\ \hline
%  induced time strip boundaries
$\tsbulkend^\pm$ & induced time strip boundaries & $d-1$ & below \req{tsbulkdef} 
\\ \hline
% bulk Residual entropy
$\REbulk$ & bulk residual entropy & $0$ &
\req{REbulkconj} and below
\\ \hline
\end{tabular}
\end{center}
\label{t:glossary}
\end{table}%

%~~~~~~~~~~~~~~~~~~~~~~~~~~~~~~~~~~~~~~~~~~~~~~~
\section{Explicit causal constructs in AdS$_3$}
\label{s:geods}
%~~~~~~~~~~~~~~~~~~~~~~~~~~~~~~~~~~~~~~~~~~

In the main body of this paper, we tried to confine our discussion to abstract reasoning, supported by illustrative examples in the form of plots.  To avoid breaking the flow of the narrative, and because it contained no conceptual subtleties, we did not present the explicit expressions we used in generating our plots.  We remedy that omission presently.

The basic constructs are bulk causal sets, and as such, they are bounded by null surfaces $\cong$ generated by null geodesics which emanate normally to the defining starting surface (namely the time strip boundaries $\tsend^\pm$ in the case of the \GCW\ $\CWg$, and bulk hole rim $\bcrv$ in the case of the \GEW\ $\EWg$).  These null geodesics, however, need not be complete: they cease to generate the causal set in question once they encounter other generators.
The essential ingredients in the explicit construction, then, are: 1) finding null geodesics in our spacetime, and 2) determining where they intersect.  The former is actually composed of two tasks, namely obtaining the general form for any null geodesic, and specifying the correct initial conditions (i.e.\ the starting position along with the constants of motion, which determines the particular geodesic of interest).
Our task is somewhat simplified by the fact that null geodesics are insensitive to conformal factor of the spacetime metric, so we can work with conformally-rescaled metric when convenient.  

Although our constructs are applicable in full generality, for illustrative purposes we have presented our plots for the example of pure AdS$_3$, where we can easily obtain the geodesics in a closed form.
Starting with the usual static coordinates for global AdS given in \req{globAdSmetr} and compactifying the radial coordinate  $r \equiv \tan \rho$, we can write the AdS$_3$ metric in the form
\begin{equation}
ds^2 
	= \frac{- dt^2 + d\rho^2 + \sin^2\! \rho \, d\ph^2}{\cos^2\! \rho}
\label{AdSrho}
\end{equation}	
with $\rho \in (0,\frac{\pi}{2})$ and $\ph \in (0, 2 \pi)$.  The boundary is at $\rho = \frac{\pi}{2}$, and lives on $R^1 \times S^1$ with metric $ds_{\rm bdy}^2 = - dt^2 +d\ph^2$.
We can now define a conformally rescaled metric by stripping off the conformal factor:
\begin{equation}
d{\tilde s}^2 \equiv ( \cos^2\! \rho ) \, ds^2 
	= - dt^2 + d\rho^2 + \sin^2\! \rho \, d\ph^2
\label{confAdSrho}
\end{equation}	
which is the metric we'll work with in what follows.
The coordinates $\{ t, \rho, \ph \}$ are precisely the ones we use for our plots, as specified in \fig{f:TS}, so that radial null geodesics are plotted at 45 degrees.

Let us denote the tangent vector along a null geodesic by
\begin{equation}
p^a = {\dot t} \, \partial_t^a 
	+ {\dot \rho} \,  \partial_\rho^a
	+ {\dot \ph} \,  \partial_\ph^a
\label{}
\end{equation}	
where $\dot{ } \equiv \frac{d}{d\lambda}$ with $\lambda$ being the affine parameter, taken to increase towards the future (independently of whether we're considering future-directed or past-directed geodesic).
Since the spacetime \req{confAdSrho} is static and rotationally symmetric, null geodesics have a conserved energy $E = -p_a \, \partial_t^a =  {\dot t}$  which we can WLOG set $= 1$ by fixing the affine parameter suitably, and angular momentum $L= p_a \, \partial_\ph^a = \sin^2\! \rho \, {\dot \ph}$.  In fact it is more convenient to work in terms of the reduced angular momentum $\ell = L/E$, which is related to the radial turning point (i.e.\ the minimum radius reached) $\rhotp$, 
\begin{equation}
\ell = \pm \sin \rhotp \ ,
\label{}
\end{equation}	
the sign depending on the sign of $\frac{d\ph}{dt}$ along the geodesic.
If we denote the corresponding temporal and angular coordinates attained at this point by $\ttp$ and $\phtp$, we can express the null geodesics compactly as
\begin{equation}
\cos(t-\ttp) = \frac{\cos \rho}{\cos \rhotp}
\qquad {\rm and} \qquad
\cos(\ph-\phtp) = \frac{\tan \rhotp}{\tan \rho}
\label{}
\end{equation}	
The turning point coordinates $(\ttp,\rhotp,\phtp)$ relate to the boundary endpoint coordinates $(\tbdy,\rhobdy,\phbdy)$ by
\begin{equation}
\tbdy = \ttp \pm \frac{\pi}{2} 
\ , \qquad 
\rhobdy =  \frac{\pi}{2} 
\ , \qquad 
\phbdy = \phtp \pm \frac{\pi}{2} 
\label{}
\end{equation}	
In fact, a slightly more convenient form for our purposes is to write the geodesics in terms of the latter, and parameterize\footnote{
Using the radial coordinate $\rho$ to  parameterize null geodesics in AdS generally involves a slight subtlety: due to the existence of the turning point $\rho = \rhotp$, we need to keep track of which $\rho > \rhotp$ branch we are on.  However, when the wedges we consider do not extend too far,  such as in all examples presented here, continuing a geodesic past its turning point is not necessary.
}
the $t$ and $\ph$ values by $\rho$,
\begin{equation}
t(\rho) = \tbdy^\pm \mp \sin^{-1} \left[ \frac{\cos \rho}{\sqrt{1-\ell^2}} \right]
\qquad {\rm and} \qquad
\ph(\rho) = \phbdy^\pm \mp \sin^{-1}\left[  \frac{\ell}{\sqrt{1-\ell^2}} \, 
 \frac{\cos \rho}{\sin \rho} \right]
\label{geodtphiofrho}
\end{equation}	
where the $\mp$ relative sign corresponds to outgoing/ingoing geodesics.
This form is most useful when we wish to identify the null normal congruence to $\tsend^\pm$, given by $\tbdy^\pm(\phbdy^\pm)$.

Alternately, if we know the initial position $\{ t_0, \rho_0 , \ph_0 \}$, such as when our starting point is the bulk curve $\bcrv$, we can rewrite \req{geodtphiofrho} in terms of this initial position by replacing 
\begin{equation}
\tbdy^\pm
= t_0 \pm  \sin^{-1} \left[ \frac{\cos \rho_0}{\sqrt{1-\ell^2}} \right]
\qquad {\rm and} \qquad
\phbdy^\pm = \ph_0 \pm \sin^{-1}\left[  \frac{\ell}{\sqrt{1-\ell^2}} \, 
 \frac{\cos \rho_0}{\sin \rho_0} \right]
\label{endptinitpt}
\end{equation}	
In either case, this leaves only the angular momentum $\ell$ to specify, corresponding to characterizing the initial velocity.  For either starting point of the residual entropy construction (the bulk hole defined by $\bcrv$, or the boundary time strip defined by $\tsend^\pm$), the requisite null normals are those emanating normally to the given curve.

Let us start with the task of ascertaining $\ell$ such that a geodesic from $\bcrv$ starts out in a perpendicular direction. 
The bulk curve $\bcrv$ can be paraterized by $\bcpar$, i.e.\ specified by the functions $\{ \tc(\bcpar) , \rhoc(\bcpar) , \phc(\bcpar) \} $, with tangent vector at each $\bcpar$ given by
\begin{equation}
k^a = \tc\dth  \, \partial_t^a 
	+ \rhoc\dth \,  \partial_\rho^a
	+ \phc\dth \,  \partial_\ph^a
\label{}
\end{equation}	
where $\dth\equiv \frac{d}{d\bcpar}$.
A null geodesic which emanates normally to $\bcrv$ then satisfies the condition $p_a \, k^a = 0$, which gives the relation
\begin{equation}
- \tc\dth \pm \sqrt{1 - \frac{\ell^2}{\sin^2 \rho}} \, \rhoc\dth + \ell \, \phc\dth = 0
\label{ellcondC}
\end{equation}	
where  $+ (-)$ sign in the second term corresponds to outgoing (ingoing) null normal.
We can easily solve \req{ellcondC} for $\ell$.
Using the further notational simplification\footnote{
Here we are implicitly assuming that $\phc(\bcpar)$ is monotonic; but if it is not, we can perform this simplification piecewise; only the radially pointing parts of $\bcrv$ give an ill-defined expression, and we can handle these separately.
} that 
$'\equiv \frac{d}{d\ph}$ along $\bcrv$ (so that 
$\tc'= \frac{d\tc}{d\phc} = \frac{\tc\dth}{\phc\dth} $ and similarly for $\phc'$), we find:
\begin{equation}
\ell = \frac{\tc' \mp \rhoc' \, \sqrt{1+\frac{\tc'^2-\rhoc'^2}{\sin^2 \rho}}}{1+\frac{\rhoc'^2}{\sin^2 \rho}}
\label{ellgenC}
\end{equation}	
(where  the $\mp$ sign corresponds to outgoing/ingoing direction; in the examples of $\EWg$ explicitly plotted here, only the $-$ sign was was relevant).

Similarly, we can find the requisite angular momentum along an ingoing null geodesic from $\tsend^\pm$ such that it emanates normally.  Here our task is even simpler, since the tangent vector $\xi^a_\pm $ to $\tsend^\pm$ is more restricted.  There is no radial component by virtue of $\tsend$ lying within the AdS boundary, and we can furthermore use $\ph = \phbdy$ to parameterize the curve since it is achronal and therefore $\tbdy(\phbdy)$ is single-valued.  This gives the tangent vector
\begin{equation}
\xi^a
 = \tbdy' \, \partial_t^a + \partial_\ph^a
\label{}
\end{equation}	
from which the requisite angular momentum is simply
\begin{equation}
\ell = \tbdy' 
\label{}
\end{equation}	
as can indeed be obtained from the expression \req{ellgenC} by setting $\rhoc'=0$ and replacing $\tc$ by $\tbdy$.
For completeness, the null normal from $\tsend^\pm$ then has the tangent vector
\begin{equation}
p^a_\pm = \partial_t^a 
	\pm \sqrt{1-\tbdy'^2} \, \partial_\rho^a
	+ \tbdy' \,  \partial_\ph^a
\label{}
\end{equation}	
(where to avoid further notational clutter we suppressed the $\pm$ on $\tbdy(\phbdy)$, which describes two separate functions, corresponding to $\tsend^+$ and $\tsend^-$).

Having the full specification of null geodesics at hand, it only remains to find their intersections.
For the \GEW\ bounded by $\cong_{\bcrv}^\pm$, this is simpler, since the generators go from $\bcrv$ to the boundary, so that we don't need to find intersections between $\cong_{\bcrv}^+$ and $\cong_{\bcrv}^-$, but only within each congruence individually.  There are then two possibilities:  either a given generator makes it all the way to the boundary without intersecting another generator, in which case the corresponding endpoint $(\rho=\frac{\pi}{2}, t = \tbdy, \ph=\phbdy) \in \tsend$, or it doesn't.  To find the separation between these two possibilities, we need to solve the pair of equations
\begin{equation}
\tbdy(\bcpar_1) = \tbdy(\bcpar_2)
\qquad {\rm and} \qquad
\phbdy(\bcpar_1) = \phbdy(\bcpar_2)
\label{causticeqns}
\end{equation}	
for the pair of unknowns $(\bcpar_1, \bcpar_2)$.
We can in fact simplify this by writing
\begin{equation}
\begin{split}
\sin \tbdy &= \frac{1}{\sqrt{1-\ell^2}} \left[ \sqrt{\sin^2 \rhoc - \ell^2}
\, \sin \tc + \cos \rhoc \, \cos \tc
\right] \\
\sin \phbdy &= \frac{1}{\sqrt{1-\ell^2} \, \sin \rhoc} \left[ \sqrt{\sin^2 \rhoc - \ell^2}
\, \sin \phc + \ell \, \cos \rhoc \, \cos \phc
\right] 
\label{}
\end{split}
\end{equation}	
though in practice it doesn't matter as we solve these equations numerically.
Similarly, we can find the remainder of the crossover seam by solving the analog of \req{causticeqns} at finite $\rho$.  
This procedure allows us to determine the \GEW\ from a bulk curve $\bcrv$.  The converse procedure of determining the \GCW\ from a boundary time strip $\ts$ proceeds analogously, except that we now have to solve for two types of intersections, those amongst generators of $\cong_{\tsend^+}$ only or  $\cong_{\tsend^-}$ only, and those between  $\cong_{\tsend^+}$ and  $\cong_{\tsend^-}$.
The time-symmetric case is much easier, as there we are absolved of the latter, and only need to evaluate the former at $t=0$.

%%%%%%%%%%%%%%%%%%%%%%%%%%%%%%%%%%%

\bibliographystyle{JHEP}
%\bibliography{ResEntBib}

\providecommand{\href}[2]{#2}\begingroup\raggedright\endgroup

\end{document}